\documentclass[aip,apl,reprint]{revtex4-1}

\usepackage{bm}
\usepackage{amsmath}
\usepackage{xcolor,soul}
\usepackage{graphicx}
\usepackage{booktabs}

\usepackage{dcolumn}
\newcolumntype{d}[1]{D{.}{.}{#1}}

\usepackage[normalem]{ulem}

\newcommand{\vcn}[1]{\hat{#1}}
\newcommand{\vcc}[1]{\boldsymbol{\mathrm{#1}}}

\newcommand{\abs}[1]{\lvert #1 \rvert}

\renewcommand{\eqref}[1]{(\ref{#1})}

\newcommand{\komma}{\, \mathrm{,}}
\newcommand{\punkt}{\, \mathrm{.}}
\newcommand{\unit}[1]{\, \mathrm{#1}}
\newcommand{\mbohr}{\, \mu_\mathrm{B}}

\renewcommand{\eqref}[1]{(\ref{#1})}

\newcommand{\eg}{e.g.,\@ }
\newcommand{\ie}{i.e.\@ }

\newcommand{\etal}{\textit{et al.\@ }}

\begin{document}

\title{Dzyaloshinskii-Moriya interaction at disordered interfaces from \emph{ab initio} theory: robustness against intermixing and tunability through dusting}

\author{Bernd Zimmermann}
\affiliation{Peter Gr\"{u}nberg Institut and Institute for Advanced Simulation, Forschungszentrum J\"{u}lich and JARA, 52425 J\"{u}lich, Germany}
\author{William Legrand}
\affiliation{Unit\'{e} Mixte de Physique, CNRS, Thales, Univ.\ Paris-Sud, Universit\'{e} Paris-Saclay, 91767 Palaiseau, France}
\author{Davide Maccariello}
\affiliation{Unit\'{e} Mixte de Physique, CNRS, Thales, Univ.\ Paris-Sud, Universit\'{e} Paris-Saclay, 91767 Palaiseau, France}
\author{Nicolas Reyren}
\affiliation{Unit\'{e} Mixte de Physique, CNRS, Thales, Univ.\ Paris-Sud, Universit\'{e} Paris-Saclay, 91767 Palaiseau, France}
\author{Vincent Cros}
\affiliation{Unit\'{e} Mixte de Physique, CNRS, Thales, Univ.\ Paris-Sud, Universit\'{e} Paris-Saclay, 91767 Palaiseau, France}
\author{Stefan Bl\"{u}gel}
\affiliation{Peter Gr\"{u}nberg Institut and Institute for Advanced Simulation, Forschungszentrum J\"{u}lich and JARA, 52425 J\"{u}lich, Germany}
\author{Albert Fert}
\affiliation{Unit\'{e} Mixte de Physique, CNRS, Thales, Univ.\ Paris-Sud, Universit\'{e} Paris-Saclay, 91767 Palaiseau, France}

\date{\today}

\begin{abstract}
The Dzyaloshinskii-Moriya interaction (DMI), which is essential for the stabilization of topologically non-trivial chiral magnetic textures such as skyrmions, is particularly strong in heterostructures of ultra-thin magnetic materials and heavy elements. We explore by density-functional theory calculations the possibility to modify the magnetic properties at Co/Pt interfaces with chemical disorder. In these systems, we find a particular robustness of the DMI against intermixing. Upon dusting the interface with a third element (all $4d$ transition metals and B, Cu, Au and Bi), a strong reduction of the DMI is predicted. This opens up possibilities to tune the DMI through the degrees of intermixing and dusting.
\end{abstract}

\maketitle

%============================
%\section{Introduction}%=====
%============================

The Dzyaloshinskii-Moriya interaction (DMI) \cite{Dzyaloshinskii1957,*Moriya1960} is an antisymmetric magnetic exchange interaction, which stabilizes chiral non-collinear magnetic states. It occurs due to spin-orbit interaction in an inversion-asymmetric crystal field. In competition with other magnetic interactions, such as magnetocrystalline anisotropy, it can lead to the occurrence of chiral domain walls \cite{Kubetzka2003,*Heide2008,*Thiaville2012}, spin spirals \cite{Bode2007,*Zimmermann2014a}, magnetic skyrmions \cite{Bogdanov1994,*Roessler2006} and anti-skyrmions \cite{Hoffmann2017}.

Magnetic multilayers or ultrathin magnetic films in contact with a substrate or a layer with strong spin-orbit interaction (SOI), such as 5$d$ metals, are important material classes where DMI arises, and then it is commonly referred to as interfacial DMI. In such systems, chiral magnetic structures have been theoretically predicted \cite{Heinze2011} and experimentally observed \cite{Moreau-Luchaire2016,Boulle2016,Romming2013,*Soumyanarayanan2017}. Especially skyrmionic magnetic objects exhibit fascinating properties: they are compact chiral solitonic objects and probably the smallest stable magnetic textures \cite{Kiselev2011}, their topological nature adds to the stability of skyrmions, and they can be efficiently moved by spin-orbit torques. The magnetic texture imposed by DMI may have also technological implications for more mature technologies such as new generations of magnetic memory, logic or neuro-inspired skyrmion based devices \cite{Fert2017:NatureMaterialsReview}.

In spite of extensive research efforts over the past years, many fundamental questions remain unresolved. Of particular interest is the relation of DMI with structural inhomogeneities, such as grain boundaries \cite{Li2017}, single atomic defects \cite{Mueller2015}, step edges, roughness and intermixing. It is evident that the magnetic properties of nano-scale objects alter significantly as a result of such inhomogeneities \cite{Fernandes:18.1}. As an example, reconstruction lines in Fe/Ir(111) lead to an alteration of the wavefront of spin spirals \cite{Finco2017} and finally to the stabilization of skyrmions that are not axially symmetric \cite{Hagemeister2016,*Finco2016}.

Another aspect that is strongly linked to structural defects is the pinning of skyrmions and domain walls. Several micromagnetic simulations show that materials with a granular structure pin skyrmions, where the exchange stiffness \cite{Leliaert2014}, magnetocrystalline anisotropy \cite{JVKim2017,*Voto2016,*Raposo2017} or DMI \cite{Legrand2017} are altered from one grain to the next. As shown by Legrand \etal\cite{Legrand2017}, the impact of pinning is strongly enhanced if the grains and skyrmions are of similar size. In line is the observation that materials with an amorphous crystal structure seem to exhibit less pinning than polycrystalline samples \cite{Woo2016}.
%This might explain the robust movement of skyrmions in amorphous Pt/CoFeB/MgO multilayers as compared to polycrystalline Pt/Co/Ta multilayers \cite{Woo2016}.
On the other hand, is has been shown that the DMI is sensitive to the interface quality: in a symmetric Pt/Co/Pt stack, the DMI from the two interfaces cancel nearly perfectly (as theoretically expected) only if the growth is epitaxial \cite{Hrabec2014}. In contrast, a finite DMI is observed if the upper and lower interfaces are of different crystallographic quality, which can be fine-tuned by sputtering-deposition conditions \cite{Wells2017}.

However, many details are still not understood on a microscopic level, \eg how such a modification of DMI could arise and how large it can be. For example, tuning the pressure during sputtering deposition may change the roughness as well as the intermixing at an interface \cite{Wells2017}, which are difficult to disentangle experimentally. Here, \emph{ab initio} calculations can unambiguously predict the impact of certain structural modification and hence guide experimentalists in their optimization of samples.

In this Letter, we study the effect of two different modifications of an interface between a magnetic material and a heavy metal on the DMI by density-functional theory calculations. At first, we introduce chemical intermixing of atoms at the interface and secondly, we chemically modify the interface by dusting with various chemical elements. We choose the prominent Co/Pt bilayer for our study, as it combines several features in favor of spintronics applications: a strong perpendicular magnetic anisotropy in thin Co-films, a strong interfacial DMI, as well as a large spin Hall effect arising from Pt.

%======================
%\section{Method}%======
%======================

\begin{figure}[t!]
  \includegraphics[width=0.49\textwidth]{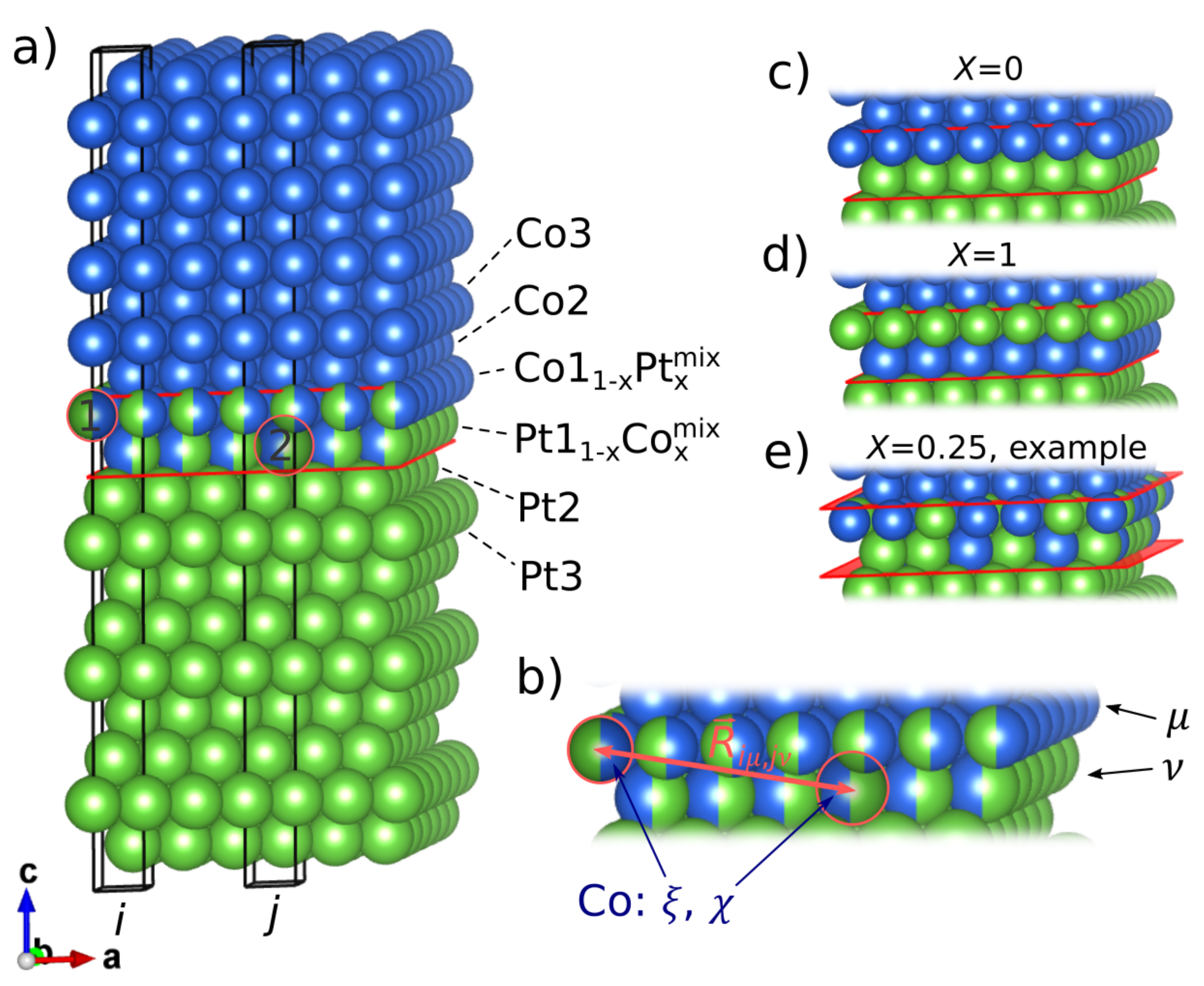}
  \caption{a): Schematic drawing of the Co/Pt bilayer. Red planes highlight the interface region where intermixing is introduced. Two different unit cells of the two-dimensional slab, labeled by $i$ and $j$, are shown. b) Zoom onto the interfacial intermixed introducing labels for the layer ($\mu$ and $\nu$), as well as for the chemical atom ($\xi$ and $\chi$) at an intermixed site. c)-e) Examples of interfaces as controlled by the intermixing parameter $x$.}
  \label{fig:unitcell}
\end{figure}

The studied systems are composed of a slab of 10 atomic layers of Co(0001) on 10 atomic layers of Pt(111), continued by two semi-infinite vacuum regions in the $\pm z$-directions. This thickness is sufficient to suppress any artifacts from the surfaces of the finite slab on properties like the DMI. We use optimized structural parameters as described in the supplementary material. We next calculate magnetic properties from density-functional theory in local density approximation \cite{VWN1980} by employing the Korringa-Kohn-Rostoker (KKR) method \cite{Bauer.phd}. We model intermixing at the interface by means of the Coherent Potential Approximation (CPA) \cite{Ebert2011} by replacing the Pt and Co interface layers by an alloy Pt$_{1-x}$Co$_x$ and Co$_{1-x}$Pt$_x$, respectively (see Fig.\ref{fig:unitcell}). Hence, increasing $x$ from zero, which corresponds to the ideal interface (see Fig.~\ref{fig:unitcell}(c)), we arrive for $x=0.5$ at the situation that 50\% of the interface Co-layer diffused into the Pt layer and vice versa (which we call ``fully intermixed interface''). Upon further increasing $x>0.5$, the interface layers tend to become more ordered but inverted, hence representing a Co(0001)/Pt/Co/Pt(111) structure for $x=1$ (see Fig.~\ref{fig:unitcell}(d)).

Moreover, in order to investigate the potential of tuning the DMI by dusting of the interface by a third chemical element, as it could be experimentally achieved \eg by co-sputtering, a second series of calculations has been performed. As dusting element $Z$ we chose all $4d$ elements as well as other prominent elements for spintronics applications, \eg Au or Bi (see results for details). To account for a realistic concentration profile, the chemical disorder is included in the first two substrate layers by replacing the Pt1 and Pt2 layers (see Fig.~\ref{fig:unitcell}(a)) with the alloy layers Pt$_{1-c} Z_c$ and Pt$_{1-c/2} Z_{c/2}$, respectively, where $c$ controls the degree of dusting.

As far as the DMI is concerned, pair-interactions are extracted with the method of infinitesimal rotations \cite{Liechtenstein1987,*Udvardi2003,*Ebert2009}, which yields parameters for an atomistic spin-lattice model, where the DMI-part for the case of chemical disorder reads
\begin{equation}
   E_\mathrm{DM} = \sum_{i,\mu,\xi} \sum_{j,\nu,\chi} \vcc{D}_{i \mu,j \nu}^{\xi,\chi} \cdot \left( c_{\mu}^{\xi} \, \vcc{S}_{i \mu}^{\xi} \times c_{\nu}^{\chi} \, \vcc{S}_{j \nu}^{\chi} \right) \punkt
   \label{eq:DMI_Heisenberg_term}
 \end{equation}
Here, $\vcc{D}_{i \mu,j \nu}^{\xi,\chi}$ is the DMI vector between two atoms which are situated at sites specified by $(i,\mu)$ and $(j,\nu)$, where the first and second index denote the unit cell and layer indices of the slab, respectively (see Fig.~\ref{fig:unitcell}(b)). The symbols $\xi$ and $\chi$ label the chemical types (\ie Co, Pt or $Z$) at these given sites with respective concentrations $c$, where $\sum_\xi c_{\mu}^{\xi} \equiv 1$. Obviously, it is not very transparent to analyze all values of such a high-dimensional space of DMI quantities, and it does also not correspond to the experimental reality where usually a scalar quantity, such as a frequency shift or a period length, is related to the DMI. Hence, we derive an effective quantity that relates the DMI to a single, experimentally measurable quantity. Such a quantity is the DMI spiralization\cite{Freimuth2013}, which can be understood as the effect of DMI on coherently rotating spin-spirals in the long-wavelength limit,
\begin{equation}
  \vcc{S}_{i \mu}^{\xi} = \mathcal{R}_{\vcn{z} \rightarrow \vcn{n}} \left( \begin{array}{c} 
          \cos(\vcc{q} \cdot (\vcc{R}_i + \vcc{\tau}_{\mu} )) \\
          \sin(\vcc{q} \cdot (\vcc{R}_i + \vcc{\tau}_{\mu} )) \\
          0
          \end{array} \right) \komma \quad \text{with} ~ \abs{\vcc{q}} \rightarrow 0 \komma
          \label{eq:ansatz_spinspiral}
 \end{equation}
 where the rotation matrix $\mathcal{R}$ turns the spin-spiral's rotation axis from its local $z$-axis into the direction $\hat{n}$, $\vcc{R}_i$ is a lattice vector and $\vcc{\tau}_{\mu}$ is the position of atom $\mu$ in the unit cell. Both, coherent rotation in all layers and long-wavelength limit are reasonable assumptions for Co-based systems because of their strong ferromagnetic nearest-neighbor exchange interaction. Inserting the ansatz \eqref{eq:ansatz_spinspiral} into the DMI part of the spin-lattice model \eqref{eq:DMI_Heisenberg_term} yields for spin-spirals along the propagation direction $\hat{e}_\gamma$ in the long-wavelength limit
\begin{equation}
   \left. \frac{\partial E_\mathrm{DM}}{\partial q_\gamma} \right\rvert_{q=0} = -N \sum_{\mu,\nu}\sum_{\xi,\chi} c_{\mu \xi} c_{\nu \chi} \, \left(\vcn{n} \cdot \underline{\underline{\mathcal{D}}}_{\mu \xi}^{\nu \chi} \cdot \vcn{e}_\gamma \right) \komma
   \label{eq:DMI_spiralization_energy}
\end{equation}
with atom-pair specific spiralization tensors
\begin{equation}
   \underline{\underline{\mathcal{D}}}_{\mu \xi}^{\nu \chi}  = \sum_{j} \vcc{D}_{0 \mu,j \nu}^{\xi,\chi} \otimes \vcc{R}_{0 \mu,j \nu} \komma
   \label{eq:DMI_spiralization_atomtypes}
\end{equation}
$N$ denotes the number of unit cells in the crystal and $\vcc{R}_{0 \mu,j \nu} = \vcc{R}_j + \vcc{\tau}_\nu - \vcc{\tau}_{\mu}$. The spiralization tensors are $3\times 3$ matrices, but due to the $C_{3v}$ symmetry of the crystal, only two components that are related by antisymmetry, $\mathcal{D}_{yx} = -\mathcal{D}_{xy}$, are non-vanishing. Hence, we succeeded in combining all the contributions of the DMI into an effective scalar quantity,
\begin{equation}
  \mathcal{D}_\mathrm{tot} = \sum_{\mu,\nu}\sum_{\xi,\chi} c_{\mu \xi} c_{\nu \chi} \, \left(\vcn{y} \cdot \underline{\underline{\mathcal{D}}}_{\mu \xi}^{\nu \chi} \cdot \vcn{x} \right) \punkt
  \label{eq:DMI_spiralization_total}
\end{equation}
We note that care has to be taken when determining the spiralization, because the sum in Eq.~\eqref{eq:DMI_spiralization_atomtypes} is difficult to converge due to a long-range oscillatory (RKKY-type) behavior of the DMI contributions in metallic samples (see supplementary material).

%======================
%\section{Results}%=====
%======================

\begin{figure}[t!]
	\includegraphics[width=0.49\textwidth]{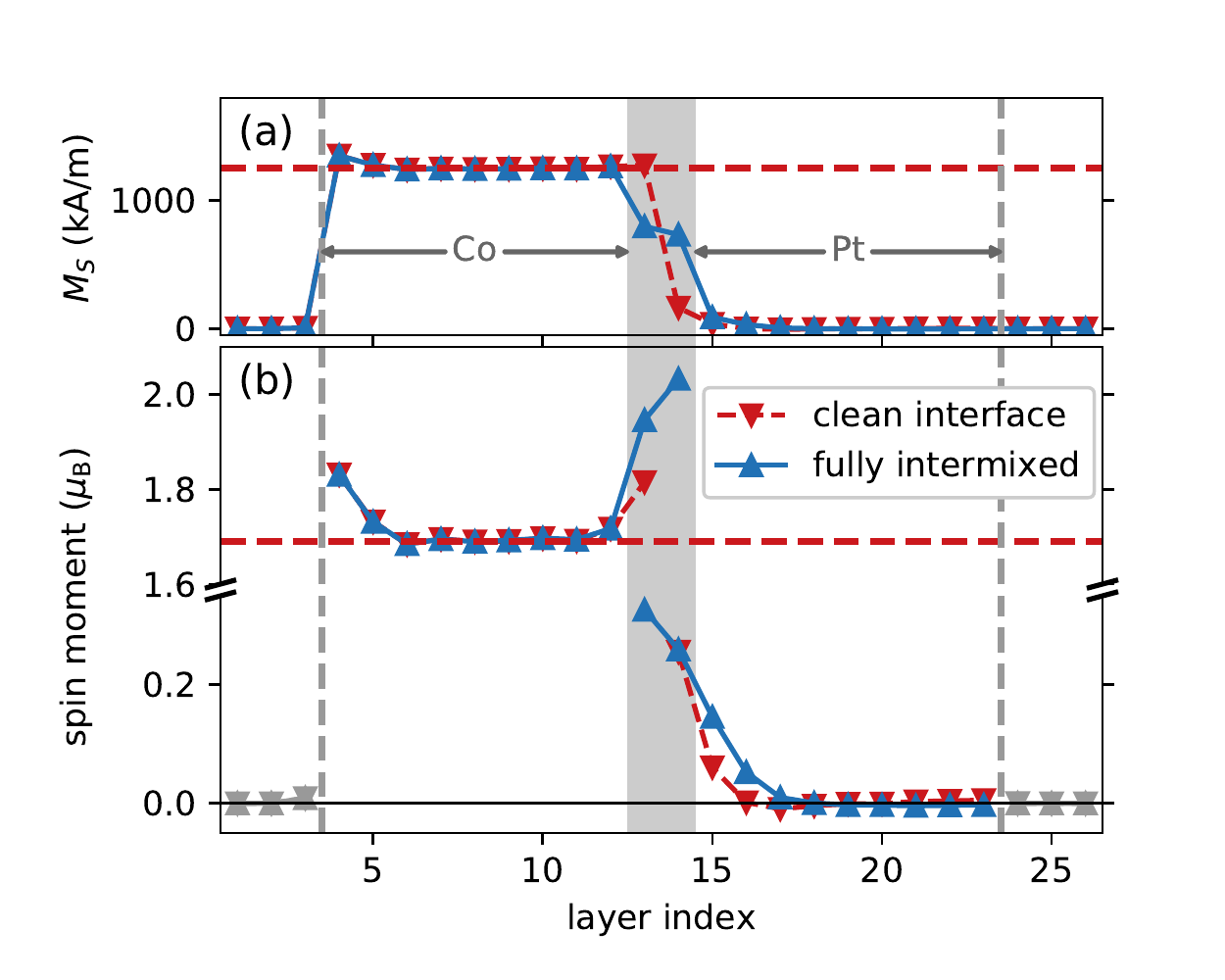}
	\caption{(a) Magnetization and (b) atomic magnetic moments across the slab for the clean ($x=0$) and completely intermixed ($x=0.5$) interfaces. The horizontal dashed lines indicate the bulk Co values of $1250 \unit{kA/m}$ and $1.69 \mbohr$, respectively. Vertical dashed lines denote the position of the top and bottom surfaces, and the interface region with possible intermixing is highlighted in gray. The layer index starts from the top vacuum, \ie the Co side.}
	\label{fig:magnetic_moments}
\end{figure}

We first discuss the magnetization and atomic magnetic moments (see Fig.~\ref{fig:magnetic_moments}) across the slab. In the interior of the Co part, the magnetic moments take the bulk-value of $1.69 \mbohr$, which corresponds to a saturation magnetization of $M_\mathrm{S}=1250 \unit{kA/m}$, being slightly enhanced at the surface and interface by 8\% and 7\%, respectively (see Fig.~\ref{fig:magnetic_moments}(b)). In the clean case, the magnetization and moments in the Pt part drop quickly and have only a sizable contribution in the first Pt interface layer ($0.26 \mbohr$). This value agrees very well with experimentally determined induced spin moment of $(0.24 \pm 0.05) \unit{\mu_\mathrm{B}}$ measured by X-ray magnetic circular dichroism (XMCD) on ID12 beamline at ESRF on Pt/Co/Ir multilayers and a recent \emph{ab initio} study of a Co monolayer on Pt(111) \cite{Belabbes}. Upon inclusion of disorder, we see essentially two modifications: Firstly, the magnetization in the intermixed region (see gray zone in Fig.\ref{fig:magnetic_moments}) naturally washes out due to Co dilution and exhibits an intermediate value of about $700 \unit{kA/m}$ for the completely intermixed interface. However, the local Co moments are even more enhanced as compared to the clean interface ($1.95 \mbohr$ in the nominal Co-layer, and $2.03 \mbohr$ for Co atoms which have been alloyed into the nominal Pt layer, see Fig.~\ref{fig:magnetic_moments}(b)). Such an increase is in line with a narrowing of the $d$ bands and promoted magnetism due to the progressive reduction of like-nearest neighbors: from 12 in the bulk, via 9 at the clean interface, until in average 7.5 and 4.5 for the Co moments at the fully intermixed interface in the nominal Co and Pt-layers, respectively.
Secondly, the induced moments in the Pt-part are also enhanced and the decay into the Pt-bulk is slightly slower as compared to the clean interface. However, the induced Pt moments only contribute little to the total magnetic moment of the slab (around 2\%), in accordance with previous studies \cite{Zarpellon2012}. All Co moments in the intermixed layers vary linearly as function of intermixing $x$ (not shown).

\begin{figure}[t!]
  \includegraphics[width=0.49\textwidth]{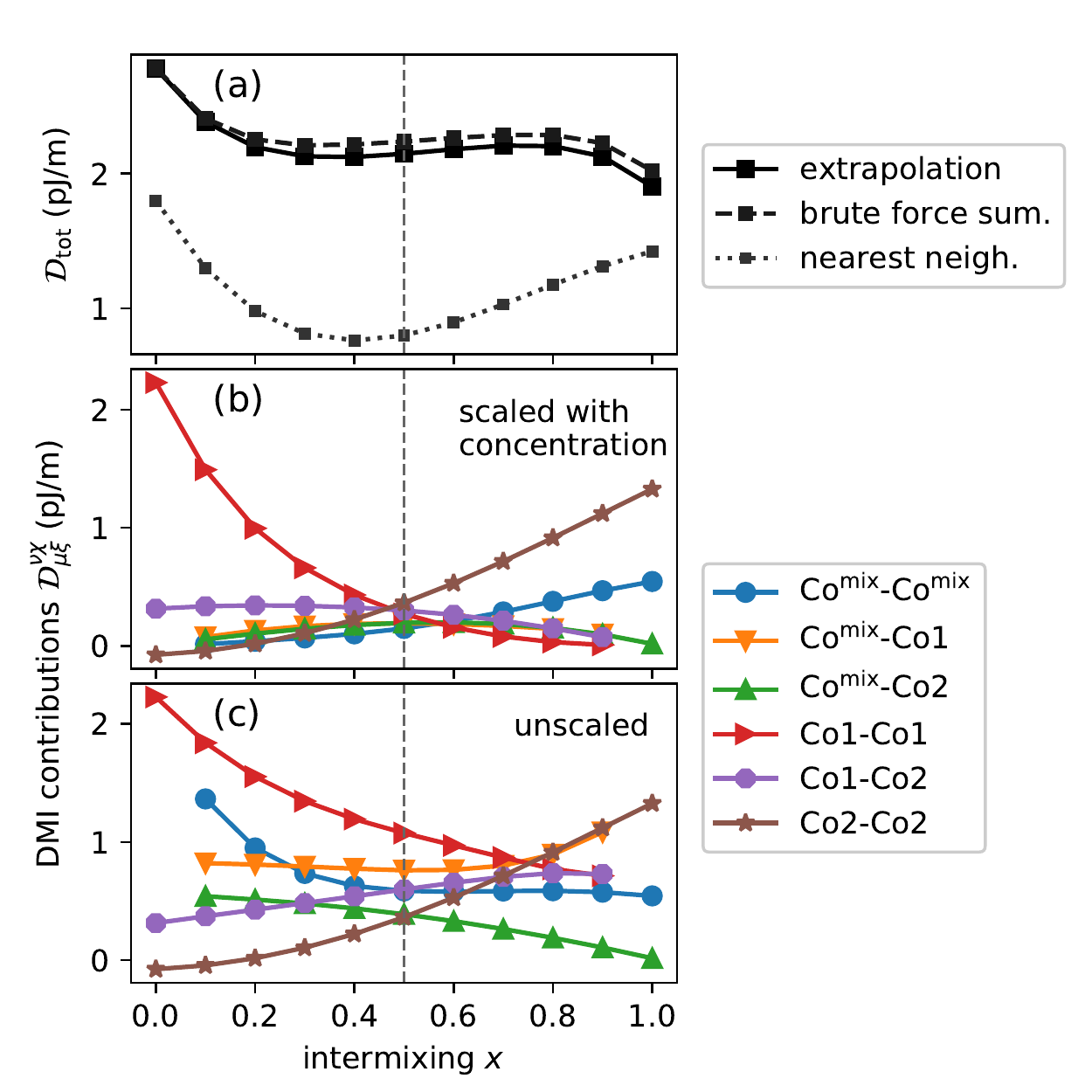}
  \caption{(a) Total DMI spiralization as function of intermixing in three different approaches for evaluating the sum over neighbors:
      sophisticated extrapolation technique (solid line), 
      brute force summation up to $6 \, a_\mathrm{lat}$ (dashed line), 
      and taking only nearest neighbors (dotted line).
      (b-c) Contributions from specific pairs of layers to the total spiralization.
      The vertical dashed line indicates the case for a fully intermixed interface ($x=0.5$).
      }
  \label{fig:DMI_intermixing_all}
\end{figure}

Next we turn to the Dzyaloshinskii-Moriya interaction. For the clean interface, we determine an interface spiralization of $\mathcal{D}_\mathrm{tot} = 2.78 \, \mathrm{pJ/m}$, being in perfect agreement to earlier calculations on a system with comparable structural setup but calulated with a different electronic structure method \cite{Freimuth2013}. The positive sign corresponds to a lowered energy of magnetic states with left-handed (anti-clockwise) chirality, \ie $\uparrow \leftarrow \downarrow$ \cite{*[{See Appendix B of }] [{ for a detailed discussion of the chirality.}] Schweflinghaus2016}. This value is somewhat larger than the experimentally determined interface DMI for the Co/Pt interface, which ranges between $1.2$ and $2.2~\mathrm{pJ/m}$ \cite{Moreau-Luchaire2016,Boulle2016,Cho2015,*Di2015,*Belmeguenai2015,*Kim2015,*Ma2018} depending on the experimental details. We note for completeness, that standard micromagnetic DMI parameters are obtained by dividing the interface spiralization by the thickness of the Co layer, yielding a DMI of $2.53~\mathrm{mJ/m^2}$ for a 1.1~nm thick Co film.

Upon inclusion of intermixing, the DMI drops by only 20\% for $x=0.2$, and remains rather constant for stronger intermixing ($0.2<x<0.8$) (see Fig.~\ref{fig:DMI_intermixing_all}(a)). For $x>0.8$, where the system tends to become more ordered again, DMI drops again by about 10\%. This very robust DMI against intermixing constitutes one of the main messages of this work and results from compensating trends for the individual pair contributions to the total spiralization: the dominant intralayer spiralization from the interface Co-layer (\ie $\vcn{y} \cdot \underline{\underline{\mathcal{D}}}_\text{Co1}^\text{Co1} \cdot \vcn{x}$, which we simply term Co1--Co1 spiralization) continuously drops with $x$ (see red curve in Fig.~\ref{fig:DMI_intermixing_all}(b)), but at the same time, the intralayer spiralization in the next Co-layer (\ie Co2--Co2) increases and yields the strongest contribution for $x>0.5$.

This trend can be intuitively understood by considering two aspects: the local inversion asymmetry and spin-orbit strength mediated by the surrounding Pt atoms. The interface Co-layer (Co1) exhibits the strongest structural inversion asymmetry for $x=0$, getting reduced for increasing intermixing (as also evident from the magnetization profile, see Fig.~\ref{fig:magnetic_moments}(a)), whereas the Pt-content in the same and adjacent layers remains constant. In contrast, both measures increase with $x$ for the Co2 layer, in line with the increasing DMI spiralization for Co2--Co2 and interlayer Co1--Co2 contributions. Interestingly, all contributions add with the same chirality and hence cause an overall large DMI (except negligible Co2--Co2 contributions for $x<0.2$).

We shortly comment on the convergence of the DMI spiralization tensor in Eq.~\eqref{eq:DMI_spiralization_atomtypes} with respect to the number of neighbors included: taking neighbours up to a distance of $6~a_\mathrm{lat}$ into account or whether alternatively applying a careful extrapolation technique yields consistent results. We find taking only nearest neighbors into account greatly fails to predict the robustness of DMI against intermixing at the Co-Pt interface (see Fig.~\ref{fig:DMI_intermixing_all}(a) and supplementary material for details).

We briefly compare to the scarce data on the influence of intermixing in the literature: Wells \etal \cite{Wells2017} controlled in symmetric $\mathrm{Pt}\vert \mathrm{Co} \vert \mathrm{Pt}$ trilayers the quality of the upper and lower interface by different annealing temperatures during the growth of the stack. They observe a net DMI on the order of $\mathcal{D} =  0.25 \, \mathrm{pJ/m}$ \footnote{We converted the reported DMI-fields of the order of $200 \, \mathrm{Oe}$ into an interface DMI using $\mathcal{D}=(\mu_0 \, H_\mathrm{DMI} \, M_S \, \sqrt{A/K_0})/t$, all values from Wells \etal \cite{Wells2017}.} and anticipate it to be mostly due to intermixing, which is compatible to the 20\% decrease found here. In another study, Lavrijsen \etal \cite{Lavrijsen2015} have noticed a certain insensitivity of DMI against growth conditions in contrast to the magnetocrystalline anisotropy, the latter one changing by a factor of 3, supporting our findings of a robustness of DMI. In an earlier theoretical work, Yang \etal \cite{Yang2014} have modeled 25\% intermixing and found a larger destructive effect of $–50\%$ on the DMI. The main advantages of our approach are the refined treatment of disorder (by means of CPA as opposed to a supercell approach used by Yang \textit{et al.}) and the utilization of the long-wavelength limit (with important effects on the electronic structure).

\begin{figure}[t!]
  \includegraphics[width=0.49\textwidth]{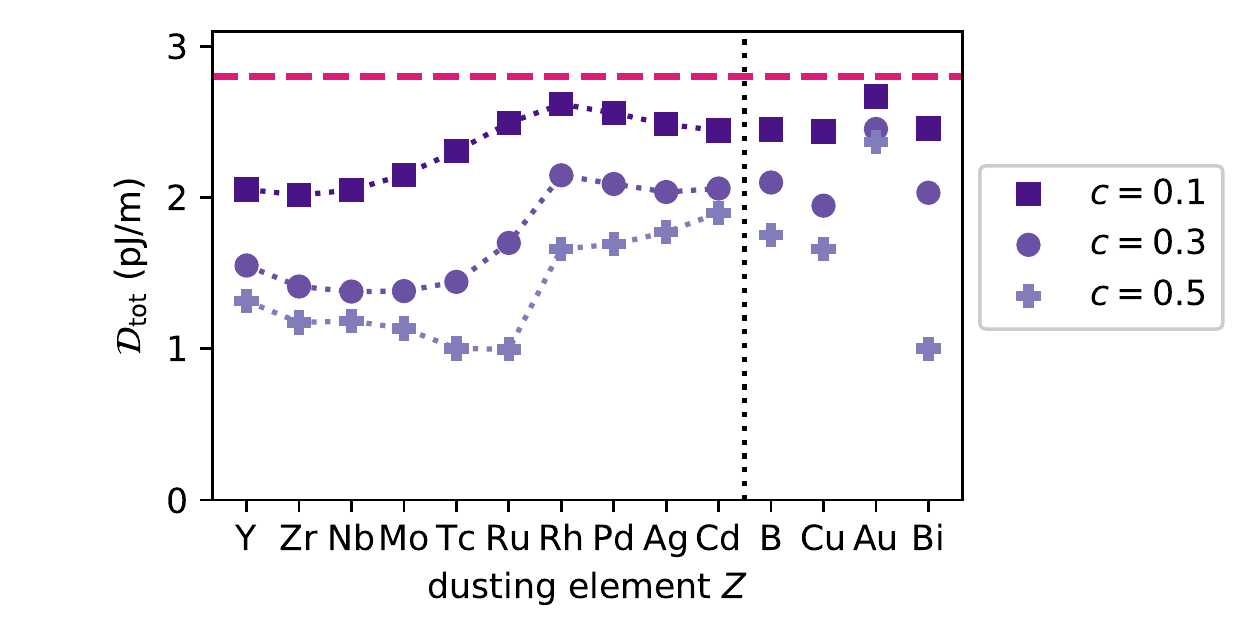}
  \caption{Total DMI $\mathcal{D}_\mathrm{tot}$ for three different dusting concentrations $c$ as function of the dusting element.}
  \label{fig:dusting}
\end{figure}

A second objective of this work is to examine the possibility to tune the DMI by dusting the substrate. We included the dusting element in the first and second Pt layers with concentrations $c$ and $c/2$, and choose $c \in \{0.1,0.3,0.5\}$. Fig.~\ref{fig:dusting} summarizes the results showing always a reducing effect of dusting on the DMI, irrespective of the dusting element $Z$ or the chosen degree of dusting (compare to $\mathcal{D}_\mathrm{tot} = 2.78 \, \mathrm{pJ/m}$ of the clean interface, as indicated by the dashed line in Fig.~\ref{fig:dusting}. A few general trends can be deduced: a stronger dusting always results in a stronger decrease of the DMI. Upon light dusting, we obtain a maximal reduction of DMI by 30\% for early $4d$ elements (Y--Nb), whereas for strong dusting, the reduction is stronger ($65\%$) and shifts towards the middle of the $4d$ series (Tc and Ru). In fact, a pronounced step in the DMI evolves in the middle of the 4d series as $c$ is increased.

A striking insensitivity of DMI is obtained for an interface which is dusted with Au: even for large doping concentrations ($c=0.5$), we obtain only a 15\% reduction. In contrast, upon dusting the interface with the similarly `heavy' element Bi reveals a huge destructive effect ($-65\%$) for $c=0.5$.

%==========================
%\section{Conclusions}%=====
%==========================

In conclusion, we have investigated the effect of intermixing and dusting of the Co/Pt interface by density-functional theory calculations. We reveal a peculiar robustness of DMI against intermixing: after an initial drop by only 20\% of the large value for the ideal Co/Pt interface, a nearly constant DMI was computed for a broad range of degrees of intermixing. We explain this as a compensation effect, where not only the Co-interface layer, but many more layers add constructively to yield a robust spiralization. Based on the electronic similarities we expect a similar behavior for Co/Ir(111). This information is important for DMI based technologies which frequently depend on Co/Pt or Co/Ir interfaces because variations in growth lead only to small changes of the DMI. Whether the DMI of other 3$d$-5$d$ transition-metal interfaces are also insensitive to intermixing is a matter of future investigations. Secondly, we demonstrate that the DMI can be modified by dusting the interface with a third chemical element. Dusting of Pt by $4d$ metals, Bi, Cu or Au, always reduces the DMI, and the reduction can be as large as $-65\%$, which may also show that Pt-produced DMI values are not easy to top. An interesting insensitivity with respect to the dusting with Au is obtained, which might be exploited to keep the strong DMI of the Co/Pt interface, but engineer some other key property for spintronics. Overall, our findings motivate experimentalists to exploit the degree of intermixing and dusting to engineer interface properties.

\bigskip

See supplementary material for structural details of the DFT calculations and additional information on the convergence of DMI spiralization.

%===============================
%\section{Acknowledgements}%====
%===============================

\bigskip

We acknowledge discussions with P.~Mavropoulos, P.~R\"{u}{\ss}mann, M.~Hoffmann, B.~Dup\'{e}, A.~Rogalev and F.~Wilhelm. B.Z.\ acknowledges CNRS/Thales for their hospitality during the course of this work, and a postdoc fellowship of the German Academic Exchange Service. We acknowledge funding from the European Union grant MAGicSky (No.\ 665095), DARPA TEE program (MIPR\# HR0011831554) from DOI, as well as computing time at the JURECA supercomputer (project jias1a04) of the J\"ulich Supercomputing Centre and JARA-HPC of RWTH Aachen University.

\bibliography{references}

\begin{thebibliography}{49}%
\makeatletter
\providecommand \@ifxundefined [1]{%
 \@ifx{#1\undefined}
}%
\providecommand \@ifnum [1]{%
 \ifnum #1\expandafter \@firstoftwo
 \else \expandafter \@secondoftwo
 \fi
}%
\providecommand \@ifx [1]{%
 \ifx #1\expandafter \@firstoftwo
 \else \expandafter \@secondoftwo
 \fi
}%
\providecommand \natexlab [1]{#1}%
\providecommand \enquote  [1]{``#1''}%
\providecommand \bibnamefont  [1]{#1}%
\providecommand \bibfnamefont [1]{#1}%
\providecommand \citenamefont [1]{#1}%
\providecommand \href@noop [0]{\@secondoftwo}%
\providecommand \href [0]{\begingroup \@sanitize@url \@href}%
\providecommand \@href[1]{\@@startlink{#1}\@@href}%
\providecommand \@@href[1]{\endgroup#1\@@endlink}%
\providecommand \@sanitize@url [0]{\catcode `\\12\catcode `\$12\catcode
  `\&12\catcode `\#12\catcode `\^12\catcode `\_12\catcode `\%12\relax}%
\providecommand \@@startlink[1]{}%
\providecommand \@@endlink[0]{}%
\providecommand \url  [0]{\begingroup\@sanitize@url \@url }%
\providecommand \@url [1]{\endgroup\@href {#1}{\urlprefix }}%
\providecommand \urlprefix  [0]{URL }%
\providecommand \Eprint [0]{\href }%
\providecommand \doibase [0]{http://dx.doi.org/}%
\providecommand \selectlanguage [0]{\@gobble}%
\providecommand \bibinfo  [0]{\@secondoftwo}%
\providecommand \bibfield  [0]{\@secondoftwo}%
\providecommand \translation [1]{[#1]}%
\providecommand \BibitemOpen [0]{}%
\providecommand \bibitemStop [0]{}%
\providecommand \bibitemNoStop [0]{.\EOS\space}%
\providecommand \EOS [0]{\spacefactor3000\relax}%
\providecommand \BibitemShut  [1]{\csname bibitem#1\endcsname}%
\let\auto@bib@innerbib\@empty
%</preamble>
\bibitem [{\citenamefont {Dzyaloshinskii}(1957)}]{Dzyaloshinskii1957}%
  \BibitemOpen
  \bibfield  {author} {\bibinfo {author} {\bibfnamefont {I.~E.}\ \bibnamefont
  {Dzyaloshinskii}},\ }\href@noop {} {\bibfield  {journal} {\bibinfo  {journal}
  {Sov.\ Phys.\ JETP}\ }\textbf {\bibinfo {volume} {5}},\ \bibinfo {pages}
  {1259} (\bibinfo {year} {1957})}\BibitemShut {NoStop}%
\bibitem [{\citenamefont {Moriya}(1960)}]{Moriya1960}%
  \BibitemOpen
  \bibfield  {author} {\bibinfo {author} {\bibfnamefont {T.}~\bibnamefont
  {Moriya}},\ }\href {\doibase 10.1103/PhysRev.120.91} {\bibfield  {journal}
  {\bibinfo  {journal} {Phys. Rev.}\ }\textbf {\bibinfo {volume} {120}},\
  \bibinfo {pages} {91} (\bibinfo {year} {1960})}\BibitemShut {NoStop}%
\bibitem [{\citenamefont {Kubetzka}\ \emph {et~al.}(2003)\citenamefont
  {Kubetzka}, \citenamefont {Pietzsch}, \citenamefont {Bode},\ and\
  \citenamefont {Wiesendanger}}]{Kubetzka2003}%
  \BibitemOpen
  \bibfield  {author} {\bibinfo {author} {\bibfnamefont {A.}~\bibnamefont
  {Kubetzka}}, \bibinfo {author} {\bibfnamefont {O.}~\bibnamefont {Pietzsch}},
  \bibinfo {author} {\bibfnamefont {M.}~\bibnamefont {Bode}}, \ and\ \bibinfo
  {author} {\bibfnamefont {R.}~\bibnamefont {Wiesendanger}},\ }\href {\doibase
  10.1103/PhysRevB.67.020401} {\bibfield  {journal} {\bibinfo  {journal} {Phys.
  Rev. B}\ }\textbf {\bibinfo {volume} {67}},\ \bibinfo {pages} {020401}
  (\bibinfo {year} {2003})}\BibitemShut {NoStop}%
\bibitem [{\citenamefont {Heide}, \citenamefont {Bihlmayer},\ and\
  \citenamefont {Bl\"ugel}(2008)}]{Heide2008}%
  \BibitemOpen
  \bibfield  {author} {\bibinfo {author} {\bibfnamefont {M.}~\bibnamefont
  {Heide}}, \bibinfo {author} {\bibfnamefont {G.}~\bibnamefont {Bihlmayer}}, \
  and\ \bibinfo {author} {\bibfnamefont {S.}~\bibnamefont {Bl\"ugel}},\ }\href
  {\doibase 10.1103/PhysRevB.78.140403} {\bibfield  {journal} {\bibinfo
  {journal} {Phys. Rev. B}\ }\textbf {\bibinfo {volume} {78}},\ \bibinfo
  {pages} {140403} (\bibinfo {year} {2008})}\BibitemShut {NoStop}%
\bibitem [{\citenamefont {Thiaville}\ \emph {et~al.}(2012)\citenamefont
  {Thiaville}, \citenamefont {Rohart}, \citenamefont {Ju\'{e}}, \citenamefont
  {Cros},\ and\ \citenamefont {Fert}}]{Thiaville2012}%
  \BibitemOpen
  \bibfield  {author} {\bibinfo {author} {\bibfnamefont {A.}~\bibnamefont
  {Thiaville}}, \bibinfo {author} {\bibfnamefont {S.}~\bibnamefont {Rohart}},
  \bibinfo {author} {\bibfnamefont {E.}~\bibnamefont {Ju\'{e}}}, \bibinfo
  {author} {\bibfnamefont {V.}~\bibnamefont {Cros}}, \ and\ \bibinfo {author}
  {\bibfnamefont {A.}~\bibnamefont {Fert}},\ }\href {\doibase
  10.1209/0295-5075/100/57002} {\bibfield  {journal} {\bibinfo  {journal}
  {EPL}\ }\textbf {\bibinfo {volume} {100}},\ \bibinfo {pages} {57002}
  (\bibinfo {year} {2012})}\BibitemShut {NoStop}%
\bibitem [{\citenamefont {Bode}\ \emph {et~al.}(2007)\citenamefont {Bode},
  \citenamefont {Heide}, \citenamefont {von Bergmann}, \citenamefont
  {Ferriani}, \citenamefont {Heinze}, \citenamefont {Bihlmayer}, \citenamefont
  {Kubetzka}, \citenamefont {Pietzsch}, \citenamefont {Bl{\"{u}}gel},\ and\
  \citenamefont {Wiesendanger}}]{Bode2007}%
  \BibitemOpen
  \bibfield  {author} {\bibinfo {author} {\bibfnamefont {M.}~\bibnamefont
  {Bode}}, \bibinfo {author} {\bibfnamefont {M.}~\bibnamefont {Heide}},
  \bibinfo {author} {\bibfnamefont {K.}~\bibnamefont {von Bergmann}}, \bibinfo
  {author} {\bibfnamefont {P.}~\bibnamefont {Ferriani}}, \bibinfo {author}
  {\bibfnamefont {S.}~\bibnamefont {Heinze}}, \bibinfo {author} {\bibfnamefont
  {G.}~\bibnamefont {Bihlmayer}}, \bibinfo {author} {\bibfnamefont
  {A.}~\bibnamefont {Kubetzka}}, \bibinfo {author} {\bibfnamefont
  {O.}~\bibnamefont {Pietzsch}}, \bibinfo {author} {\bibfnamefont
  {S.}~\bibnamefont {Bl{\"{u}}gel}}, \ and\ \bibinfo {author} {\bibfnamefont
  {R.}~\bibnamefont {Wiesendanger}},\ }\href {\doibase 10.1038/nature05802}
  {\bibfield  {journal} {\bibinfo  {journal} {Nature}\ }\textbf {\bibinfo
  {volume} {447}},\ \bibinfo {pages} {190} (\bibinfo {year}
  {2007})}\BibitemShut {NoStop}%
\bibitem [{\citenamefont {Zimmermann}\ \emph {et~al.}(2014)\citenamefont
  {Zimmermann}, \citenamefont {Heide}, \citenamefont {Bihlmayer},\ and\
  \citenamefont {Bl{\"{u}}gel}}]{Zimmermann2014a}%
  \BibitemOpen
  \bibfield  {author} {\bibinfo {author} {\bibfnamefont {B.}~\bibnamefont
  {Zimmermann}}, \bibinfo {author} {\bibfnamefont {M.}~\bibnamefont {Heide}},
  \bibinfo {author} {\bibfnamefont {G.}~\bibnamefont {Bihlmayer}}, \ and\
  \bibinfo {author} {\bibfnamefont {S.}~\bibnamefont {Bl{\"{u}}gel}},\ }\href
  {\doibase 10.1103/PhysRevB.90.115427} {\bibfield  {journal} {\bibinfo
  {journal} {Phys. Rev. B}\ }\textbf {\bibinfo {volume} {90}},\ \bibinfo
  {pages} {115427} (\bibinfo {year} {2014})}\BibitemShut {NoStop}%
\bibitem [{\citenamefont {Bogdanov}\ and\ \citenamefont
  {Hubert}(1994)}]{Bogdanov1994}%
  \BibitemOpen
  \bibfield  {author} {\bibinfo {author} {\bibfnamefont {A.}~\bibnamefont
  {Bogdanov}}\ and\ \bibinfo {author} {\bibfnamefont {A.}~\bibnamefont
  {Hubert}},\ }\href {\doibase https://doi.org/10.1016/0304-8853(94)90046-9}
  {\bibfield  {journal} {\bibinfo  {journal} {J. Magn. Magn. Mater.}\ }\textbf
  {\bibinfo {volume} {138}},\ \bibinfo {pages} {255 } (\bibinfo {year}
  {1994})}\BibitemShut {NoStop}%
\bibitem [{\citenamefont {R{\"{o}}{\ss}ler}, \citenamefont {Bogdanov},\ and\
  \citenamefont {Pfleiderer}(2006)}]{Roessler2006}%
  \BibitemOpen
  \bibfield  {author} {\bibinfo {author} {\bibfnamefont {U.~K.}\ \bibnamefont
  {R{\"{o}}{\ss}ler}}, \bibinfo {author} {\bibfnamefont {A.~N.}\ \bibnamefont
  {Bogdanov}}, \ and\ \bibinfo {author} {\bibfnamefont {C.}~\bibnamefont
  {Pfleiderer}},\ }\href {\doibase 10.1038/nature05056} {\bibfield  {journal}
  {\bibinfo  {journal} {Nature}\ }\textbf {\bibinfo {volume} {442}},\ \bibinfo
  {pages} {797} (\bibinfo {year} {2006})}\BibitemShut {NoStop}%
\bibitem [{\citenamefont {Hoffmann}\ \emph {et~al.}(2017)\citenamefont
  {Hoffmann}, \citenamefont {Zimmermann}, \citenamefont {M\"{u}ller},
  \citenamefont {Sch\"{u}rhoff}, \citenamefont {Kiselev}, \citenamefont
  {Melcher},\ and\ \citenamefont {Bl\"{u}gel}}]{Hoffmann2017}%
  \BibitemOpen
  \bibfield  {author} {\bibinfo {author} {\bibfnamefont {M.}~\bibnamefont
  {Hoffmann}}, \bibinfo {author} {\bibfnamefont {B.}~\bibnamefont
  {Zimmermann}}, \bibinfo {author} {\bibfnamefont {G.~P.}\ \bibnamefont
  {M\"{u}ller}}, \bibinfo {author} {\bibfnamefont {D.}~\bibnamefont
  {Sch\"{u}rhoff}}, \bibinfo {author} {\bibfnamefont {N.~S.}\ \bibnamefont
  {Kiselev}}, \bibinfo {author} {\bibfnamefont {C.}~\bibnamefont {Melcher}}, \
  and\ \bibinfo {author} {\bibfnamefont {S.}~\bibnamefont {Bl\"{u}gel}},\
  }\href {\doibase 10.1038/s41467-017-00313-0} {\bibfield  {journal} {\bibinfo
  {journal} {Nat. Commun.}\ }\textbf {\bibinfo {volume} {8}},\ \bibinfo {pages}
  {308} (\bibinfo {year} {2017})}\BibitemShut {NoStop}%
\bibitem [{\citenamefont {Heinze}\ \emph {et~al.}(2011)\citenamefont {Heinze},
  \citenamefont {{Von Bergmann}}, \citenamefont {Menzel}, \citenamefont
  {Brede}, \citenamefont {Kubetzka}, \citenamefont {Wiesendanger},
  \citenamefont {Bihlmayer},\ and\ \citenamefont {Bl{\"{u}}gel}}]{Heinze2011}%
  \BibitemOpen
  \bibfield  {author} {\bibinfo {author} {\bibfnamefont {S.}~\bibnamefont
  {Heinze}}, \bibinfo {author} {\bibfnamefont {K.}~\bibnamefont {{Von
  Bergmann}}}, \bibinfo {author} {\bibfnamefont {M.}~\bibnamefont {Menzel}},
  \bibinfo {author} {\bibfnamefont {J.}~\bibnamefont {Brede}}, \bibinfo
  {author} {\bibfnamefont {A.}~\bibnamefont {Kubetzka}}, \bibinfo {author}
  {\bibfnamefont {R.}~\bibnamefont {Wiesendanger}}, \bibinfo {author}
  {\bibfnamefont {G.}~\bibnamefont {Bihlmayer}}, \ and\ \bibinfo {author}
  {\bibfnamefont {S.}~\bibnamefont {Bl{\"{u}}gel}},\ }\href {\doibase
  10.1038/nphys2045} {\bibfield  {journal} {\bibinfo  {journal} {Nat. Phys.}\
  }\textbf {\bibinfo {volume} {7}},\ \bibinfo {pages} {713} (\bibinfo {year}
  {2011})}\BibitemShut {NoStop}%
\bibitem [{\citenamefont {Moreau-Luchaire}\ \emph {et~al.}(2016)\citenamefont
  {Moreau-Luchaire}, \citenamefont {Moutafis}, \citenamefont {Reyren},
  \citenamefont {Sampaio}, \citenamefont {Vaz}, \citenamefont {{Van Horne}},
  \citenamefont {Bouzehouane}, \citenamefont {Garcia}, \citenamefont
  {Deranlot}, \citenamefont {Warnicke}, \citenamefont {Wohlh{\"{u}}ter},
  \citenamefont {George}, \citenamefont {Weigand}, \citenamefont {Raabe},
  \citenamefont {Cros},\ and\ \citenamefont {Fert}}]{Moreau-Luchaire2016}%
  \BibitemOpen
  \bibfield  {author} {\bibinfo {author} {\bibfnamefont {C.}~\bibnamefont
  {Moreau-Luchaire}}, \bibinfo {author} {\bibfnamefont {C.}~\bibnamefont
  {Moutafis}}, \bibinfo {author} {\bibfnamefont {N.}~\bibnamefont {Reyren}},
  \bibinfo {author} {\bibfnamefont {J.}~\bibnamefont {Sampaio}}, \bibinfo
  {author} {\bibfnamefont {C.~A.~F.}\ \bibnamefont {Vaz}}, \bibinfo {author}
  {\bibfnamefont {N.}~\bibnamefont {{Van Horne}}}, \bibinfo {author}
  {\bibfnamefont {K.}~\bibnamefont {Bouzehouane}}, \bibinfo {author}
  {\bibfnamefont {K.}~\bibnamefont {Garcia}}, \bibinfo {author} {\bibfnamefont
  {C.}~\bibnamefont {Deranlot}}, \bibinfo {author} {\bibfnamefont
  {P.}~\bibnamefont {Warnicke}}, \bibinfo {author} {\bibfnamefont
  {P.}~\bibnamefont {Wohlh{\"{u}}ter}}, \bibinfo {author} {\bibfnamefont
  {J.-M.}\ \bibnamefont {George}}, \bibinfo {author} {\bibfnamefont
  {M.}~\bibnamefont {Weigand}}, \bibinfo {author} {\bibfnamefont
  {J.}~\bibnamefont {Raabe}}, \bibinfo {author} {\bibfnamefont
  {V.}~\bibnamefont {Cros}}, \ and\ \bibinfo {author} {\bibfnamefont
  {A.}~\bibnamefont {Fert}},\ }\href {\doibase 10.1038/nnano.2015.313}
  {\bibfield  {journal} {\bibinfo  {journal} {Nat. Nanotechnol.}\ }\textbf
  {\bibinfo {volume} {11}},\ \bibinfo {pages} {444} (\bibinfo {year}
  {2016})}\BibitemShut {NoStop}%
\bibitem [{\citenamefont {Boulle}\ \emph {et~al.}(2016)\citenamefont {Boulle},
  \citenamefont {Vogel}, \citenamefont {Yang}, \citenamefont {Pizzini},
  \citenamefont {{de Souza Chaves}}, \citenamefont {Locatelli}, \citenamefont
  {Menteş}, \citenamefont {Sala}, \citenamefont {Buda-Prejbeanu},
  \citenamefont {Klein}, \citenamefont {Belmeguenai}, \citenamefont
  {Roussign{\'{e}}}, \citenamefont {Stashkevich}, \citenamefont {Ch{\'{e}}rif},
  \citenamefont {Aballe}, \citenamefont {Foerster}, \citenamefont {Chshiev},
  \citenamefont {Auffret}, \citenamefont {Miron},\ and\ \citenamefont
  {Gaudin}}]{Boulle2016}%
  \BibitemOpen
  \bibfield  {author} {\bibinfo {author} {\bibfnamefont {O.}~\bibnamefont
  {Boulle}}, \bibinfo {author} {\bibfnamefont {J.}~\bibnamefont {Vogel}},
  \bibinfo {author} {\bibfnamefont {H.}~\bibnamefont {Yang}}, \bibinfo {author}
  {\bibfnamefont {S.}~\bibnamefont {Pizzini}}, \bibinfo {author} {\bibfnamefont
  {D.}~\bibnamefont {{de Souza Chaves}}}, \bibinfo {author} {\bibfnamefont
  {A.}~\bibnamefont {Locatelli}}, \bibinfo {author} {\bibfnamefont {T.~O.}\
  \bibnamefont {Menteş}}, \bibinfo {author} {\bibfnamefont {A.}~\bibnamefont
  {Sala}}, \bibinfo {author} {\bibfnamefont {L.~D.}\ \bibnamefont
  {Buda-Prejbeanu}}, \bibinfo {author} {\bibfnamefont {O.}~\bibnamefont
  {Klein}}, \bibinfo {author} {\bibfnamefont {M.}~\bibnamefont {Belmeguenai}},
  \bibinfo {author} {\bibfnamefont {Y.}~\bibnamefont {Roussign{\'{e}}}},
  \bibinfo {author} {\bibfnamefont {A.}~\bibnamefont {Stashkevich}}, \bibinfo
  {author} {\bibfnamefont {S.~M.}\ \bibnamefont {Ch{\'{e}}rif}}, \bibinfo
  {author} {\bibfnamefont {L.}~\bibnamefont {Aballe}}, \bibinfo {author}
  {\bibfnamefont {M.}~\bibnamefont {Foerster}}, \bibinfo {author}
  {\bibfnamefont {M.}~\bibnamefont {Chshiev}}, \bibinfo {author} {\bibfnamefont
  {S.}~\bibnamefont {Auffret}}, \bibinfo {author} {\bibfnamefont {I.~M.}\
  \bibnamefont {Miron}}, \ and\ \bibinfo {author} {\bibfnamefont
  {G.}~\bibnamefont {Gaudin}},\ }\href
  {http://dx.doi.org/10.1038/nnano.2015.315} {\bibfield  {journal} {\bibinfo
  {journal} {Nat. Nanotechnol.}\ }\textbf {\bibinfo {volume} {11}},\ \bibinfo
  {pages} {449} (\bibinfo {year} {2016})}\BibitemShut {NoStop}%
\bibitem [{\citenamefont {Romming}\ \emph {et~al.}(2013)\citenamefont
  {Romming}, \citenamefont {Hanneken}, \citenamefont {Menzel}, \citenamefont
  {Bickel}, \citenamefont {Wolter}, \citenamefont {von Bergmann}, \citenamefont
  {Kubetzka},\ and\ \citenamefont {Wiesendanger}}]{Romming2013}%
  \BibitemOpen
  \bibfield  {author} {\bibinfo {author} {\bibfnamefont {N.}~\bibnamefont
  {Romming}}, \bibinfo {author} {\bibfnamefont {C.}~\bibnamefont {Hanneken}},
  \bibinfo {author} {\bibfnamefont {M.}~\bibnamefont {Menzel}}, \bibinfo
  {author} {\bibfnamefont {J.~E.}\ \bibnamefont {Bickel}}, \bibinfo {author}
  {\bibfnamefont {B.}~\bibnamefont {Wolter}}, \bibinfo {author} {\bibfnamefont
  {K.}~\bibnamefont {von Bergmann}}, \bibinfo {author} {\bibfnamefont
  {A.}~\bibnamefont {Kubetzka}}, \ and\ \bibinfo {author} {\bibfnamefont
  {R.}~\bibnamefont {Wiesendanger}},\ }\href {\doibase 10.1126/science.1240573}
  {\bibfield  {journal} {\bibinfo  {journal} {Science}\ }\textbf {\bibinfo
  {volume} {341}},\ \bibinfo {pages} {636} (\bibinfo {year}
  {2013})}\BibitemShut {NoStop}%
\bibitem [{\citenamefont {Soumyanarayanan}\ \emph {et~al.}(2017)\citenamefont
  {Soumyanarayanan}, \citenamefont {Raju}, \citenamefont {{Gonzalez Oyarce}},
  \citenamefont {Tan}, \citenamefont {Im}, \citenamefont {Petrovi{\'{c}}},
  \citenamefont {Ho}, \citenamefont {Khoo}, \citenamefont {Tran}, \citenamefont
  {Gan}, \citenamefont {Ernult},\ and\ \citenamefont
  {Panagopoulos}}]{Soumyanarayanan2017}%
  \BibitemOpen
  \bibfield  {author} {\bibinfo {author} {\bibfnamefont {A.}~\bibnamefont
  {Soumyanarayanan}}, \bibinfo {author} {\bibfnamefont {M.}~\bibnamefont
  {Raju}}, \bibinfo {author} {\bibfnamefont {A.~L.}\ \bibnamefont {{Gonzalez
  Oyarce}}}, \bibinfo {author} {\bibfnamefont {A.~K.~C.}\ \bibnamefont {Tan}},
  \bibinfo {author} {\bibfnamefont {M.-Y.}\ \bibnamefont {Im}}, \bibinfo
  {author} {\bibfnamefont {A.}~\bibnamefont {Petrovi{\'{c}}}}, \bibinfo
  {author} {\bibfnamefont {P.}~\bibnamefont {Ho}}, \bibinfo {author}
  {\bibfnamefont {K.~H.}\ \bibnamefont {Khoo}}, \bibinfo {author}
  {\bibfnamefont {M.}~\bibnamefont {Tran}}, \bibinfo {author} {\bibfnamefont
  {C.~K.}\ \bibnamefont {Gan}}, \bibinfo {author} {\bibfnamefont
  {F.}~\bibnamefont {Ernult}}, \ and\ \bibinfo {author} {\bibfnamefont
  {C.}~\bibnamefont {Panagopoulos}},\ }\href
  {http://dx.doi.org/10.1038/nmat4934} {\bibfield  {journal} {\bibinfo
  {journal} {Nat. Mater.}\ }\textbf {\bibinfo {volume} {16}},\ \bibinfo {pages}
  {898} (\bibinfo {year} {2017})}\BibitemShut {NoStop}%
\bibitem [{\citenamefont {Kiselev}\ \emph {et~al.}(2011)\citenamefont
  {Kiselev}, \citenamefont {Bogdanov}, \citenamefont {Sch{\"{a}}fer},\ and\
  \citenamefont {R{\"{o}}ler}}]{Kiselev2011}%
  \BibitemOpen
  \bibfield  {author} {\bibinfo {author} {\bibfnamefont {N.~S.}\ \bibnamefont
  {Kiselev}}, \bibinfo {author} {\bibfnamefont {A.~N.}\ \bibnamefont
  {Bogdanov}}, \bibinfo {author} {\bibfnamefont {R.}~\bibnamefont
  {Sch{\"{a}}fer}}, \ and\ \bibinfo {author} {\bibfnamefont {U.~K.}\
  \bibnamefont {R{\"{o}}ler}},\ }\href {\doibase
  10.1088/0022-3727/44/39/392001} {\bibfield  {journal} {\bibinfo  {journal}
  {J. Phys. D}\ }\textbf {\bibinfo {volume} {44}},\ \bibinfo {pages} {392001}
  (\bibinfo {year} {2011})}\BibitemShut {NoStop}%
\bibitem [{\citenamefont {Fert}, \citenamefont {Reyren},\ and\ \citenamefont
  {Cros}(2017)}]{Fert2017:NatureMaterialsReview}%
  \BibitemOpen
  \bibfield  {author} {\bibinfo {author} {\bibfnamefont {A.}~\bibnamefont
  {Fert}}, \bibinfo {author} {\bibfnamefont {N.}~\bibnamefont {Reyren}}, \ and\
  \bibinfo {author} {\bibfnamefont {V.}~\bibnamefont {Cros}},\ }\href {\doibase
  10.1038/natrevmats.2017.31} {\bibfield  {journal} {\bibinfo  {journal} {Nat.
  Rev. Mater.}\ }\textbf {\bibinfo {volume} {2}},\ \bibinfo {pages} {17031}
  (\bibinfo {year} {2017})}\BibitemShut {NoStop}%
\bibitem [{\citenamefont {Li}\ \emph {et~al.}(2017)\citenamefont {Li},
  \citenamefont {Zheng}, \citenamefont {Tavabi}, \citenamefont {Caron},
  \citenamefont {Jin}, \citenamefont {Du}, \citenamefont {Kovács},
  \citenamefont {Tian}, \citenamefont {Farle},\ and\ \citenamefont
  {Dunin-Borkowski}}]{Li2017}%
  \BibitemOpen
  \bibfield  {author} {\bibinfo {author} {\bibfnamefont {Z.-A.}\ \bibnamefont
  {Li}}, \bibinfo {author} {\bibfnamefont {F.}~\bibnamefont {Zheng}}, \bibinfo
  {author} {\bibfnamefont {A.~H.}\ \bibnamefont {Tavabi}}, \bibinfo {author}
  {\bibfnamefont {J.}~\bibnamefont {Caron}}, \bibinfo {author} {\bibfnamefont
  {C.}~\bibnamefont {Jin}}, \bibinfo {author} {\bibfnamefont {H.}~\bibnamefont
  {Du}}, \bibinfo {author} {\bibfnamefont {A.}~\bibnamefont {Kovács}},
  \bibinfo {author} {\bibfnamefont {M.}~\bibnamefont {Tian}}, \bibinfo {author}
  {\bibfnamefont {M.}~\bibnamefont {Farle}}, \ and\ \bibinfo {author}
  {\bibfnamefont {R.~E.}\ \bibnamefont {Dunin-Borkowski}},\ }\href {\doibase
  10.1021/acs.nanolett.6b04280} {\bibfield  {journal} {\bibinfo  {journal}
  {Nano Lett.}\ }\textbf {\bibinfo {volume} {17}},\ \bibinfo {pages} {1395}
  (\bibinfo {year} {2017})}\BibitemShut {NoStop}%
\bibitem [{\citenamefont {M\"uller}\ and\ \citenamefont
  {Rosch}(2015)}]{Mueller2015}%
  \BibitemOpen
  \bibfield  {author} {\bibinfo {author} {\bibfnamefont {J.}~\bibnamefont
  {M\"uller}}\ and\ \bibinfo {author} {\bibfnamefont {A.}~\bibnamefont
  {Rosch}},\ }\href {\doibase 10.1103/PhysRevB.91.054410} {\bibfield  {journal}
  {\bibinfo  {journal} {Phys. Rev. B}\ }\textbf {\bibinfo {volume} {91}},\
  \bibinfo {pages} {054410} (\bibinfo {year} {2015})}\BibitemShut {NoStop}%
\bibitem [{\citenamefont {Fernandes}\ \emph {et~al.}(2018)\citenamefont
  {Fernandes}, \citenamefont {Bouaziz}, \citenamefont {Bouhassoune},
  \citenamefont {Bl\"ugel},\ and\ \citenamefont {Lounis}}]{Fernandes:18.1}%
  \BibitemOpen
  \bibfield  {author} {\bibinfo {author} {\bibfnamefont {I.~L.}\ \bibnamefont
  {Fernandes}}, \bibinfo {author} {\bibfnamefont {J.}~\bibnamefont {Bouaziz}},
  \bibinfo {author} {\bibfnamefont {M.}~\bibnamefont {Bouhassoune}}, \bibinfo
  {author} {\bibfnamefont {S.}~\bibnamefont {Bl\"ugel}}, \ and\ \bibinfo
  {author} {\bibfnamefont {S.}~\bibnamefont {Lounis}},\ }\href@noop {}
  {\bibfield  {journal} {\bibinfo  {journal} {submitted to Nat. Commun.}\ }
  (\bibinfo {year} {2018})}\BibitemShut {NoStop}%
\bibitem [{\citenamefont {Finco}\ \emph {et~al.}(2017)\citenamefont {Finco},
  \citenamefont {R\'ozsa}, \citenamefont {Hsu}, \citenamefont {Kubetzka},
  \citenamefont {Vedmedenko}, \citenamefont {von Bergmann},\ and\ \citenamefont
  {Wiesendanger}}]{Finco2017}%
  \BibitemOpen
  \bibfield  {author} {\bibinfo {author} {\bibfnamefont {A.}~\bibnamefont
  {Finco}}, \bibinfo {author} {\bibfnamefont {L.}~\bibnamefont {R\'ozsa}},
  \bibinfo {author} {\bibfnamefont {P.-J.}\ \bibnamefont {Hsu}}, \bibinfo
  {author} {\bibfnamefont {A.}~\bibnamefont {Kubetzka}}, \bibinfo {author}
  {\bibfnamefont {E.}~\bibnamefont {Vedmedenko}}, \bibinfo {author}
  {\bibfnamefont {K.}~\bibnamefont {von Bergmann}}, \ and\ \bibinfo {author}
  {\bibfnamefont {R.}~\bibnamefont {Wiesendanger}},\ }\href {\doibase
  10.1103/PhysRevLett.119.037202} {\bibfield  {journal} {\bibinfo  {journal}
  {Phys. Rev. Lett.}\ }\textbf {\bibinfo {volume} {119}},\ \bibinfo {pages}
  {037202} (\bibinfo {year} {2017})}\BibitemShut {NoStop}%
\bibitem [{\citenamefont {Hagemeister}, \citenamefont {Vedmedenko},\ and\
  \citenamefont {Wiesendanger}(2016)}]{Hagemeister2016}%
  \BibitemOpen
  \bibfield  {author} {\bibinfo {author} {\bibfnamefont {J.}~\bibnamefont
  {Hagemeister}}, \bibinfo {author} {\bibfnamefont {E.~Y.}\ \bibnamefont
  {Vedmedenko}}, \ and\ \bibinfo {author} {\bibfnamefont {R.}~\bibnamefont
  {Wiesendanger}},\ }\href {\doibase 10.1103/PhysRevB.94.104434} {\bibfield
  {journal} {\bibinfo  {journal} {Phys. Rev. B}\ }\textbf {\bibinfo {volume}
  {94}},\ \bibinfo {pages} {104434} (\bibinfo {year} {2016})}\BibitemShut
  {NoStop}%
\bibitem [{\citenamefont {Finco}\ \emph {et~al.}(2016)\citenamefont {Finco},
  \citenamefont {Hsu}, \citenamefont {Kubetzka}, \citenamefont {von Bergmann},\
  and\ \citenamefont {Wiesendanger}}]{Finco2016}%
  \BibitemOpen
  \bibfield  {author} {\bibinfo {author} {\bibfnamefont {A.}~\bibnamefont
  {Finco}}, \bibinfo {author} {\bibfnamefont {P.-J.}\ \bibnamefont {Hsu}},
  \bibinfo {author} {\bibfnamefont {A.}~\bibnamefont {Kubetzka}}, \bibinfo
  {author} {\bibfnamefont {K.}~\bibnamefont {von Bergmann}}, \ and\ \bibinfo
  {author} {\bibfnamefont {R.}~\bibnamefont {Wiesendanger}},\ }\href {\doibase
  10.1103/PhysRevB.94.214402} {\bibfield  {journal} {\bibinfo  {journal} {Phys.
  Rev. B}\ }\textbf {\bibinfo {volume} {94}},\ \bibinfo {pages} {214402}
  (\bibinfo {year} {2016})}\BibitemShut {NoStop}%
\bibitem [{\citenamefont {Leliaert}\ \emph {et~al.}(2014)\citenamefont
  {Leliaert}, \citenamefont {{Van De Wiele}}, \citenamefont {Vansteenkiste},
  \citenamefont {Laurson}, \citenamefont {Durin}, \citenamefont {Dupr{\'{e}}},\
  and\ \citenamefont {{Van Waeyenberge}}}]{Leliaert2014}%
  \BibitemOpen
  \bibfield  {author} {\bibinfo {author} {\bibfnamefont {J.}~\bibnamefont
  {Leliaert}}, \bibinfo {author} {\bibfnamefont {B.}~\bibnamefont {{Van De
  Wiele}}}, \bibinfo {author} {\bibfnamefont {A.}~\bibnamefont
  {Vansteenkiste}}, \bibinfo {author} {\bibfnamefont {L.}~\bibnamefont
  {Laurson}}, \bibinfo {author} {\bibfnamefont {G.}~\bibnamefont {Durin}},
  \bibinfo {author} {\bibfnamefont {L.}~\bibnamefont {Dupr{\'{e}}}}, \ and\
  \bibinfo {author} {\bibfnamefont {B.}~\bibnamefont {{Van Waeyenberge}}},\
  }\href {\doibase 10.1063/1.4854956} {\bibfield  {journal} {\bibinfo
  {journal} {J. Appl. Phys.}\ }\textbf {\bibinfo {volume} {115}},\ \bibinfo
  {pages} {17D102} (\bibinfo {year} {2014})}\BibitemShut {NoStop}%
\bibitem [{\citenamefont {Kim}\ and\ \citenamefont {Yoo}(2017)}]{JVKim2017}%
  \BibitemOpen
  \bibfield  {author} {\bibinfo {author} {\bibfnamefont {J.~V.}\ \bibnamefont
  {Kim}}\ and\ \bibinfo {author} {\bibfnamefont {M.~W.}\ \bibnamefont {Yoo}},\
  }\href {\doibase 10.1063/1.4979316} {\bibfield  {journal} {\bibinfo
  {journal} {Appl. Phys. Lett.}\ }\textbf {\bibinfo {volume} {110}},\ \bibinfo
  {pages} {132404} (\bibinfo {year} {2017})}\BibitemShut {NoStop}%
\bibitem [{\citenamefont {Voto}, \citenamefont {Lopez-Diaz},\ and\
  \citenamefont {Torres}(2016)}]{Voto2016}%
  \BibitemOpen
  \bibfield  {author} {\bibinfo {author} {\bibfnamefont {M.}~\bibnamefont
  {Voto}}, \bibinfo {author} {\bibfnamefont {L.}~\bibnamefont {Lopez-Diaz}}, \
  and\ \bibinfo {author} {\bibfnamefont {L.}~\bibnamefont {Torres}},\ }\href
  {\doibase 10.1088/0022-3727/49/18/185001} {\bibfield  {journal} {\bibinfo
  {journal} {J. Phys. D}\ }\textbf {\bibinfo {volume} {49}},\ \bibinfo {pages}
  {185001} (\bibinfo {year} {2016})}\BibitemShut {NoStop}%
\bibitem [{\citenamefont {Raposo}, \citenamefont {{Luis Martinez}},\ and\
  \citenamefont {Martinez}(2017)}]{Raposo2017}%
  \BibitemOpen
  \bibfield  {author} {\bibinfo {author} {\bibfnamefont {V.}~\bibnamefont
  {Raposo}}, \bibinfo {author} {\bibfnamefont {R.~F.}\ \bibnamefont {{Luis
  Martinez}}}, \ and\ \bibinfo {author} {\bibfnamefont {E.}~\bibnamefont
  {Martinez}},\ }\href {\doibase 10.1063/1.4975658} {\bibfield  {journal}
  {\bibinfo  {journal} {AIP Adv.}\ }\textbf {\bibinfo {volume} {7}},\ \bibinfo
  {pages} {056017} (\bibinfo {year} {2017})}\BibitemShut {NoStop}%
\bibitem [{\citenamefont {Legrand}\ \emph {et~al.}(2017)\citenamefont
  {Legrand}, \citenamefont {Maccariello}, \citenamefont {Reyren}, \citenamefont
  {Garcia}, \citenamefont {Moutafis}, \citenamefont {Moreau-Luchaire},
  \citenamefont {Collin}, \citenamefont {Bouzehouane}, \citenamefont {Cros},\
  and\ \citenamefont {Fert}}]{Legrand2017}%
  \BibitemOpen
  \bibfield  {author} {\bibinfo {author} {\bibfnamefont {W.}~\bibnamefont
  {Legrand}}, \bibinfo {author} {\bibfnamefont {D.}~\bibnamefont
  {Maccariello}}, \bibinfo {author} {\bibfnamefont {N.}~\bibnamefont {Reyren}},
  \bibinfo {author} {\bibfnamefont {K.}~\bibnamefont {Garcia}}, \bibinfo
  {author} {\bibfnamefont {C.}~\bibnamefont {Moutafis}}, \bibinfo {author}
  {\bibfnamefont {C.}~\bibnamefont {Moreau-Luchaire}}, \bibinfo {author}
  {\bibfnamefont {S.}~\bibnamefont {Collin}}, \bibinfo {author} {\bibfnamefont
  {K.}~\bibnamefont {Bouzehouane}}, \bibinfo {author} {\bibfnamefont
  {V.}~\bibnamefont {Cros}}, \ and\ \bibinfo {author} {\bibfnamefont
  {A.}~\bibnamefont {Fert}},\ }\href {\doibase 10.1021/acs.nanolett.7b00649}
  {\bibfield  {journal} {\bibinfo  {journal} {Nano Lett.}\ }\textbf {\bibinfo
  {volume} {17}},\ \bibinfo {pages} {2703} (\bibinfo {year}
  {2017})}\BibitemShut {NoStop}%
\bibitem [{\citenamefont {Woo}\ \emph {et~al.}(2016)\citenamefont {Woo},
  \citenamefont {Litzius}, \citenamefont {Kr{\"{u}}ger}, \citenamefont {Im},
  \citenamefont {Caretta}, \citenamefont {Richter}, \citenamefont {Mann},
  \citenamefont {Krone}, \citenamefont {Reeve}, \citenamefont {Weigand},
  \citenamefont {Agrawal}, \citenamefont {Lemesh}, \citenamefont {Mawass},
  \citenamefont {Fischer}, \citenamefont {Kl{\"{a}}ui},\ and\ \citenamefont
  {Beach}}]{Woo2016}%
  \BibitemOpen
  \bibfield  {author} {\bibinfo {author} {\bibfnamefont {S.}~\bibnamefont
  {Woo}}, \bibinfo {author} {\bibfnamefont {K.}~\bibnamefont {Litzius}},
  \bibinfo {author} {\bibfnamefont {B.}~\bibnamefont {Kr{\"{u}}ger}}, \bibinfo
  {author} {\bibfnamefont {M.-Y.}\ \bibnamefont {Im}}, \bibinfo {author}
  {\bibfnamefont {L.}~\bibnamefont {Caretta}}, \bibinfo {author} {\bibfnamefont
  {K.}~\bibnamefont {Richter}}, \bibinfo {author} {\bibfnamefont
  {M.}~\bibnamefont {Mann}}, \bibinfo {author} {\bibfnamefont {A.}~\bibnamefont
  {Krone}}, \bibinfo {author} {\bibfnamefont {R.~M.}\ \bibnamefont {Reeve}},
  \bibinfo {author} {\bibfnamefont {M.}~\bibnamefont {Weigand}}, \bibinfo
  {author} {\bibfnamefont {P.}~\bibnamefont {Agrawal}}, \bibinfo {author}
  {\bibfnamefont {I.}~\bibnamefont {Lemesh}}, \bibinfo {author} {\bibfnamefont
  {M.-A.}\ \bibnamefont {Mawass}}, \bibinfo {author} {\bibfnamefont
  {P.}~\bibnamefont {Fischer}}, \bibinfo {author} {\bibfnamefont
  {M.}~\bibnamefont {Kl{\"{a}}ui}}, \ and\ \bibinfo {author} {\bibfnamefont
  {G.~S.~D.}\ \bibnamefont {Beach}},\ }\href {\doibase 10.1038/nmat4593}
  {\bibfield  {journal} {\bibinfo  {journal} {Nat. Mater.}\ }\textbf {\bibinfo
  {volume} {15}},\ \bibinfo {pages} {501} (\bibinfo {year} {2016})}\BibitemShut
  {NoStop}%
\bibitem [{\citenamefont {Hrabec}\ \emph {et~al.}(2014)\citenamefont {Hrabec},
  \citenamefont {Porter}, \citenamefont {Wells}, \citenamefont {Benitez},
  \citenamefont {Burnell}, \citenamefont {McVitie}, \citenamefont {McGrouther},
  \citenamefont {Moore},\ and\ \citenamefont {Marrows}}]{Hrabec2014}%
  \BibitemOpen
  \bibfield  {author} {\bibinfo {author} {\bibfnamefont {A.}~\bibnamefont
  {Hrabec}}, \bibinfo {author} {\bibfnamefont {N.~A.}\ \bibnamefont {Porter}},
  \bibinfo {author} {\bibfnamefont {A.}~\bibnamefont {Wells}}, \bibinfo
  {author} {\bibfnamefont {M.~J.}\ \bibnamefont {Benitez}}, \bibinfo {author}
  {\bibfnamefont {G.}~\bibnamefont {Burnell}}, \bibinfo {author} {\bibfnamefont
  {S.}~\bibnamefont {McVitie}}, \bibinfo {author} {\bibfnamefont
  {D.}~\bibnamefont {McGrouther}}, \bibinfo {author} {\bibfnamefont {T.~A.}\
  \bibnamefont {Moore}}, \ and\ \bibinfo {author} {\bibfnamefont {C.~H.}\
  \bibnamefont {Marrows}},\ }\href {\doibase 10.1103/PhysRevB.90.020402}
  {\bibfield  {journal} {\bibinfo  {journal} {Phys. Rev. B}\ }\textbf {\bibinfo
  {volume} {90}},\ \bibinfo {pages} {020402} (\bibinfo {year}
  {2014})}\BibitemShut {NoStop}%
\bibitem [{\citenamefont {Wells}\ \emph {et~al.}(2017)\citenamefont {Wells},
  \citenamefont {Shepley}, \citenamefont {Marrows},\ and\ \citenamefont
  {Moore}}]{Wells2017}%
  \BibitemOpen
  \bibfield  {author} {\bibinfo {author} {\bibfnamefont {A.~W.~J.}\
  \bibnamefont {Wells}}, \bibinfo {author} {\bibfnamefont {P.~M.}\ \bibnamefont
  {Shepley}}, \bibinfo {author} {\bibfnamefont {C.~H.}\ \bibnamefont
  {Marrows}}, \ and\ \bibinfo {author} {\bibfnamefont {T.~A.}\ \bibnamefont
  {Moore}},\ }\href {\doibase 10.1103/PhysRevB.95.054428} {\bibfield  {journal}
  {\bibinfo  {journal} {Phys. Rev. B}\ }\textbf {\bibinfo {volume} {95}},\
  \bibinfo {pages} {054428} (\bibinfo {year} {2017})}\BibitemShut {NoStop}%
\bibitem [{\citenamefont {Vosko}, \citenamefont {Wilk},\ and\ \citenamefont
  {Nusair}(1980)}]{VWN1980}%
  \BibitemOpen
  \bibfield  {author} {\bibinfo {author} {\bibfnamefont {S.~H.}\ \bibnamefont
  {Vosko}}, \bibinfo {author} {\bibfnamefont {L.}~\bibnamefont {Wilk}}, \ and\
  \bibinfo {author} {\bibfnamefont {M.}~\bibnamefont {Nusair}},\ }\href
  {\doibase 10.1139/p80-159} {\bibfield  {journal} {\bibinfo  {journal} {Can.
  J. Phys.}\ }\textbf {\bibinfo {volume} {58}},\ \bibinfo {pages} {1200}
  (\bibinfo {year} {1980})}\BibitemShut {NoStop}%
\bibitem [{\citenamefont {Bauer}(2013)}]{Bauer.phd}%
  \BibitemOpen
  \bibfield  {author} {\bibinfo {author} {\bibfnamefont {D.~S.~G.}\
  \bibnamefont {Bauer}},\ }\emph {\bibinfo {title} {Development of a
  relativistic full-potential first-principles multiple scattering Green
  function method applied to complex magnetic textures of nano structures at
  surfaces}},\ \href
  {http://darwin.bth.rwth-aachen.de/opus3/volltexte/2014/4925} {Ph.D. thesis},\
  \bibinfo  {school} {RWTH Aachen} (\bibinfo {year} {2013})\BibitemShut
  {NoStop}%
\bibitem [{\citenamefont {Ebert}, \citenamefont {K{\"{o}}dderitzsch},\ and\
  \citenamefont {Min{\'{a}}r}(2011)}]{Ebert2011}%
  \BibitemOpen
  \bibfield  {author} {\bibinfo {author} {\bibfnamefont {H.}~\bibnamefont
  {Ebert}}, \bibinfo {author} {\bibfnamefont {D.}~\bibnamefont
  {K{\"{o}}dderitzsch}}, \ and\ \bibinfo {author} {\bibfnamefont
  {J.}~\bibnamefont {Min{\'{a}}r}},\ }\href {\doibase
  10.1088/0034-4885/74/9/096501} {\bibfield  {journal} {\bibinfo  {journal}
  {Rep. Prog. Phys.}\ }\textbf {\bibinfo {volume} {74}},\ \bibinfo {pages}
  {096501} (\bibinfo {year} {2011})}\BibitemShut {NoStop}%
\bibitem [{\citenamefont {Liechtenstein}\ \emph {et~al.}(1987)\citenamefont
  {Liechtenstein}, \citenamefont {Katsnelson}, \citenamefont {Antropov},\ and\
  \citenamefont {Gubanov}}]{Liechtenstein1987}%
  \BibitemOpen
  \bibfield  {author} {\bibinfo {author} {\bibfnamefont {A.~I.}\ \bibnamefont
  {Liechtenstein}}, \bibinfo {author} {\bibfnamefont {M.~I.}\ \bibnamefont
  {Katsnelson}}, \bibinfo {author} {\bibfnamefont {V.~P.}\ \bibnamefont
  {Antropov}}, \ and\ \bibinfo {author} {\bibfnamefont {V.~A.}\ \bibnamefont
  {Gubanov}},\ }\href {\doibase 10.1016/0304-8853(87)90721-9} {\bibfield
  {journal} {\bibinfo  {journal} {J. Magn. Magn. Mater.}\ }\textbf {\bibinfo
  {volume} {67}},\ \bibinfo {pages} {65} (\bibinfo {year} {1987})}\BibitemShut
  {NoStop}%
\bibitem [{\citenamefont {Udvardi}\ \emph {et~al.}(2003)\citenamefont
  {Udvardi}, \citenamefont {Szunyogh}, \citenamefont {Palot{\'{a}}s},\ and\
  \citenamefont {Weinberger}}]{Udvardi2003}%
  \BibitemOpen
  \bibfield  {author} {\bibinfo {author} {\bibfnamefont {L.}~\bibnamefont
  {Udvardi}}, \bibinfo {author} {\bibfnamefont {L.}~\bibnamefont {Szunyogh}},
  \bibinfo {author} {\bibfnamefont {K.}~\bibnamefont {Palot{\'{a}}s}}, \ and\
  \bibinfo {author} {\bibfnamefont {P.}~\bibnamefont {Weinberger}},\ }\href
  {\doibase 10.1103/PhysRevB.68.104436} {\bibfield  {journal} {\bibinfo
  {journal} {Phys. Rev. B}\ }\textbf {\bibinfo {volume} {68}},\ \bibinfo
  {pages} {104436} (\bibinfo {year} {2003})}\BibitemShut {NoStop}%
\bibitem [{\citenamefont {Ebert}\ and\ \citenamefont
  {Mankovsky}(2009)}]{Ebert2009}%
  \BibitemOpen
  \bibfield  {author} {\bibinfo {author} {\bibfnamefont {H.}~\bibnamefont
  {Ebert}}\ and\ \bibinfo {author} {\bibfnamefont {S.}~\bibnamefont
  {Mankovsky}},\ }\href {\doibase 10.1103/PhysRevB.79.045209} {\bibfield
  {journal} {\bibinfo  {journal} {Phys. Rev. B}\ }\textbf {\bibinfo {volume}
  {79}},\ \bibinfo {pages} {045209} (\bibinfo {year} {2009})}\BibitemShut
  {NoStop}%
\bibitem [{\citenamefont {Freimuth}, \citenamefont {Bl\"ugel},\ and\
  \citenamefont {Mokrousov}(2014)}]{Freimuth2013}%
  \BibitemOpen
  \bibfield  {author} {\bibinfo {author} {\bibfnamefont {F.}~\bibnamefont
  {Freimuth}}, \bibinfo {author} {\bibfnamefont {S.}~\bibnamefont {Bl\"ugel}},
  \ and\ \bibinfo {author} {\bibfnamefont {Y.}~\bibnamefont {Mokrousov}},\
  }\href {http://stacks.iop.org/0953-8984/26/i=10/a=104202} {\bibfield
  {journal} {\bibinfo  {journal} {J. Phys.: Cond. Mat.}\ }\textbf {\bibinfo
  {volume} {26}},\ \bibinfo {pages} {104202} (\bibinfo {year}
  {2014})}\BibitemShut {NoStop}%
\bibitem [{\citenamefont {Belabbes}\ \emph {et~al.}(2016)\citenamefont
  {Belabbes}, \citenamefont {Bihlmayer}, \citenamefont {Bechstedt},
  \citenamefont {Bl\"ugel},\ and\ \citenamefont {Manchon}}]{Belabbes}%
  \BibitemOpen
  \bibfield  {author} {\bibinfo {author} {\bibfnamefont {A.}~\bibnamefont
  {Belabbes}}, \bibinfo {author} {\bibfnamefont {G.}~\bibnamefont {Bihlmayer}},
  \bibinfo {author} {\bibfnamefont {F.}~\bibnamefont {Bechstedt}}, \bibinfo
  {author} {\bibfnamefont {S.}~\bibnamefont {Bl\"ugel}}, \ and\ \bibinfo
  {author} {\bibfnamefont {A.}~\bibnamefont {Manchon}},\ }\href {\doibase
  10.1103/PhysRevLett.117.247202} {\bibfield  {journal} {\bibinfo  {journal}
  {Phys. Rev. Lett.}\ }\textbf {\bibinfo {volume} {117}},\ \bibinfo {pages}
  {247202} (\bibinfo {year} {2016})}\BibitemShut {NoStop}%
\bibitem [{\citenamefont {Zarpellon}\ \emph {et~al.}(2012)\citenamefont
  {Zarpellon}, \citenamefont {Jaffr\`es}, \citenamefont {Frougier},
  \citenamefont {Deranlot}, \citenamefont {George}, \citenamefont {Mosca},
  \citenamefont {Lema\^{\i}tre}, \citenamefont {Freimuth}, \citenamefont
  {Duong}, \citenamefont {Renucci},\ and\ \citenamefont
  {Marie}}]{Zarpellon2012}%
  \BibitemOpen
  \bibfield  {author} {\bibinfo {author} {\bibfnamefont {J.}~\bibnamefont
  {Zarpellon}}, \bibinfo {author} {\bibfnamefont {H.}~\bibnamefont
  {Jaffr\`es}}, \bibinfo {author} {\bibfnamefont {J.}~\bibnamefont {Frougier}},
  \bibinfo {author} {\bibfnamefont {C.}~\bibnamefont {Deranlot}}, \bibinfo
  {author} {\bibfnamefont {J.~M.}\ \bibnamefont {George}}, \bibinfo {author}
  {\bibfnamefont {D.~H.}\ \bibnamefont {Mosca}}, \bibinfo {author}
  {\bibfnamefont {A.}~\bibnamefont {Lema\^{\i}tre}}, \bibinfo {author}
  {\bibfnamefont {F.}~\bibnamefont {Freimuth}}, \bibinfo {author}
  {\bibfnamefont {Q.~H.}\ \bibnamefont {Duong}}, \bibinfo {author}
  {\bibfnamefont {P.}~\bibnamefont {Renucci}}, \ and\ \bibinfo {author}
  {\bibfnamefont {X.}~\bibnamefont {Marie}},\ }\href {\doibase
  10.1103/PhysRevB.86.205314} {\bibfield  {journal} {\bibinfo  {journal} {Phys.
  Rev. B}\ }\textbf {\bibinfo {volume} {86}},\ \bibinfo {pages} {205314}
  (\bibinfo {year} {2012})}\BibitemShut {NoStop}%
\bibitem [{\citenamefont {Schweflinghaus}\ \emph {et~al.}(2016)\citenamefont
  {Schweflinghaus}, \citenamefont {Zimmermann}, \citenamefont {Heide},
  \citenamefont {Bihlmayer},\ and\ \citenamefont
  {Bl\"ugel}}]{Schweflinghaus2016}%
  \BibitemOpen
  \bibfield  {author} {\bibinfo {author} {\bibfnamefont {B.}~\bibnamefont
  {Schweflinghaus}}, \bibinfo {author} {\bibfnamefont {B.}~\bibnamefont
  {Zimmermann}}, \bibinfo {author} {\bibfnamefont {M.}~\bibnamefont {Heide}},
  \bibinfo {author} {\bibfnamefont {G.}~\bibnamefont {Bihlmayer}}, \ and\
  \bibinfo {author} {\bibfnamefont {S.}~\bibnamefont {Bl\"ugel}},\ }\href
  {\doibase 10.1103/PhysRevB.94.024403} {\bibfield  {journal} {\bibinfo
  {journal} {Phys. Rev. B}\ }\textbf {\bibinfo {volume} {94}},\ \bibinfo
  {pages} {024403} (\bibinfo {year} {2016})}\BibitemShut {NoStop}%
\bibitem [{\citenamefont {Cho}\ \emph {et~al.}(2015)\citenamefont {Cho},
  \citenamefont {Kim}, \citenamefont {Lee}, \citenamefont {Kim}, \citenamefont
  {Lavrijsen}, \citenamefont {Solignac}, \citenamefont {Yin}, \citenamefont
  {Han}, \citenamefont {{Van Hoof}}, \citenamefont {Swagten}, \citenamefont
  {Koopmans},\ and\ \citenamefont {You}}]{Cho2015}%
  \BibitemOpen
  \bibfield  {author} {\bibinfo {author} {\bibfnamefont {J.}~\bibnamefont
  {Cho}}, \bibinfo {author} {\bibfnamefont {N.~H.}\ \bibnamefont {Kim}},
  \bibinfo {author} {\bibfnamefont {S.}~\bibnamefont {Lee}}, \bibinfo {author}
  {\bibfnamefont {J.~S.}\ \bibnamefont {Kim}}, \bibinfo {author} {\bibfnamefont
  {R.}~\bibnamefont {Lavrijsen}}, \bibinfo {author} {\bibfnamefont
  {A.}~\bibnamefont {Solignac}}, \bibinfo {author} {\bibfnamefont
  {Y.}~\bibnamefont {Yin}}, \bibinfo {author} {\bibfnamefont {D.~S.}\
  \bibnamefont {Han}}, \bibinfo {author} {\bibfnamefont {N.~J.}\ \bibnamefont
  {{Van Hoof}}}, \bibinfo {author} {\bibfnamefont {H.~J.}\ \bibnamefont
  {Swagten}}, \bibinfo {author} {\bibfnamefont {B.}~\bibnamefont {Koopmans}}, \
  and\ \bibinfo {author} {\bibfnamefont {C.~Y.}\ \bibnamefont {You}},\ }\href
  {\doibase 10.1038/ncomms8635} {\bibfield  {journal} {\bibinfo  {journal}
  {Nat. Commun.}\ }\textbf {\bibinfo {volume} {6}},\ \bibinfo {pages} {7635}
  (\bibinfo {year} {2015})}\BibitemShut {NoStop}%
\bibitem [{\citenamefont {Di}\ \emph {et~al.}(2015)\citenamefont {Di},
  \citenamefont {Zhang}, \citenamefont {Lim}, \citenamefont {Ng}, \citenamefont
  {Kuok}, \citenamefont {Yu}, \citenamefont {Yoon}, \citenamefont {Qiu},\ and\
  \citenamefont {Yang}}]{Di2015}%
  \BibitemOpen
  \bibfield  {author} {\bibinfo {author} {\bibfnamefont {K.}~\bibnamefont
  {Di}}, \bibinfo {author} {\bibfnamefont {V.~L.}\ \bibnamefont {Zhang}},
  \bibinfo {author} {\bibfnamefont {H.~S.}\ \bibnamefont {Lim}}, \bibinfo
  {author} {\bibfnamefont {S.~C.}\ \bibnamefont {Ng}}, \bibinfo {author}
  {\bibfnamefont {M.~H.}\ \bibnamefont {Kuok}}, \bibinfo {author}
  {\bibfnamefont {J.}~\bibnamefont {Yu}}, \bibinfo {author} {\bibfnamefont
  {J.}~\bibnamefont {Yoon}}, \bibinfo {author} {\bibfnamefont {X.}~\bibnamefont
  {Qiu}}, \ and\ \bibinfo {author} {\bibfnamefont {H.}~\bibnamefont {Yang}},\
  }\href {\doibase 10.1103/PhysRevLett.114.047201} {\bibfield  {journal}
  {\bibinfo  {journal} {Phys. Rev. Lett.}\ }\textbf {\bibinfo {volume} {114}},\
  \bibinfo {pages} {047201} (\bibinfo {year} {2015})}\BibitemShut {NoStop}%
\bibitem [{\citenamefont {Belmeguenai}\ \emph {et~al.}(2015)\citenamefont
  {Belmeguenai}, \citenamefont {Adam}, \citenamefont {Roussign\'e},
  \citenamefont {Eimer}, \citenamefont {Devolder}, \citenamefont {Kim},
  \citenamefont {Cherif}, \citenamefont {Stashkevich},\ and\ \citenamefont
  {Thiaville}}]{Belmeguenai2015}%
  \BibitemOpen
  \bibfield  {author} {\bibinfo {author} {\bibfnamefont {M.}~\bibnamefont
  {Belmeguenai}}, \bibinfo {author} {\bibfnamefont {J.-P.}\ \bibnamefont
  {Adam}}, \bibinfo {author} {\bibfnamefont {Y.}~\bibnamefont {Roussign\'e}},
  \bibinfo {author} {\bibfnamefont {S.}~\bibnamefont {Eimer}}, \bibinfo
  {author} {\bibfnamefont {T.}~\bibnamefont {Devolder}}, \bibinfo {author}
  {\bibfnamefont {J.-V.}\ \bibnamefont {Kim}}, \bibinfo {author} {\bibfnamefont
  {S.~M.}\ \bibnamefont {Cherif}}, \bibinfo {author} {\bibfnamefont
  {A.}~\bibnamefont {Stashkevich}}, \ and\ \bibinfo {author} {\bibfnamefont
  {A.}~\bibnamefont {Thiaville}},\ }\href {\doibase 10.1103/PhysRevB.91.180405}
  {\bibfield  {journal} {\bibinfo  {journal} {Phys. Rev. B}\ }\textbf {\bibinfo
  {volume} {91}},\ \bibinfo {pages} {180405} (\bibinfo {year}
  {2015})}\BibitemShut {NoStop}%
\bibitem [{\citenamefont {Kim}\ \emph {et~al.}(2015)\citenamefont {Kim},
  \citenamefont {Han}, \citenamefont {Jung}, \citenamefont {Cho}, \citenamefont
  {Kim}, \citenamefont {Swagten},\ and\ \citenamefont {You}}]{Kim2015}%
  \BibitemOpen
  \bibfield  {author} {\bibinfo {author} {\bibfnamefont {N.~H.}\ \bibnamefont
  {Kim}}, \bibinfo {author} {\bibfnamefont {D.~S.}\ \bibnamefont {Han}},
  \bibinfo {author} {\bibfnamefont {J.}~\bibnamefont {Jung}}, \bibinfo {author}
  {\bibfnamefont {J.}~\bibnamefont {Cho}}, \bibinfo {author} {\bibfnamefont
  {J.~S.}\ \bibnamefont {Kim}}, \bibinfo {author} {\bibfnamefont {H.~J.}\
  \bibnamefont {Swagten}}, \ and\ \bibinfo {author} {\bibfnamefont {C.~Y.}\
  \bibnamefont {You}},\ }\href {\doibase 10.1063/1.4932550} {\bibfield
  {journal} {\bibinfo  {journal} {Appl. Phys. Lett.}\ }\textbf {\bibinfo
  {volume} {107}},\ \bibinfo {pages} {142408} (\bibinfo {year}
  {2015})}\BibitemShut {NoStop}%
\bibitem [{\citenamefont {Ma}\ \emph {et~al.}(2018)\citenamefont {Ma},
  \citenamefont {Yu}, \citenamefont {Tang}, \citenamefont {Li}, \citenamefont
  {He}, \citenamefont {Shi}, \citenamefont {Wang},\ and\ \citenamefont
  {Li}}]{Ma2018}%
  \BibitemOpen
  \bibfield  {author} {\bibinfo {author} {\bibfnamefont {X.}~\bibnamefont
  {Ma}}, \bibinfo {author} {\bibfnamefont {G.}~\bibnamefont {Yu}}, \bibinfo
  {author} {\bibfnamefont {C.}~\bibnamefont {Tang}}, \bibinfo {author}
  {\bibfnamefont {X.}~\bibnamefont {Li}}, \bibinfo {author} {\bibfnamefont
  {C.}~\bibnamefont {He}}, \bibinfo {author} {\bibfnamefont {J.}~\bibnamefont
  {Shi}}, \bibinfo {author} {\bibfnamefont {K.~L.}\ \bibnamefont {Wang}}, \
  and\ \bibinfo {author} {\bibfnamefont {X.}~\bibnamefont {Li}},\ }\href
  {\doibase 10.1103/PhysRevLett.120.157204} {\bibfield  {journal} {\bibinfo
  {journal} {Phys. Rev. Lett.}\ }\textbf {\bibinfo {volume} {120}},\ \bibinfo
  {pages} {157204} (\bibinfo {year} {2018})}\BibitemShut {NoStop}%
\bibitem [{Note1()}]{Note1}%
  \BibitemOpen
  \bibinfo {note} {We converted the reported DMI-fields of the order of $200
  \protect \tmspace +\thinmuskip {.1667em} \protect \mathrm {Oe}$ into an
  interface DMI using $\protect \mathcal {D}=(\mu _0 \protect \tmspace
  +\thinmuskip {.1667em} H_\protect \mathrm {DMI} \protect \tmspace
  +\thinmuskip {.1667em} M_S \protect \tmspace +\thinmuskip {.1667em} \protect
  \sqrt {A/K_0})/t$, all values from Wells \protect \textit {et al.\spacefactor
  \@m {} }\cite {Wells2017}.}\BibitemShut {Stop}%
\bibitem [{\citenamefont {Lavrijsen}\ \emph {et~al.}(2015)\citenamefont
  {Lavrijsen}, \citenamefont {Hartmann}, \citenamefont {{Van Den Brink}},
  \citenamefont {Yin}, \citenamefont {Barcones}, \citenamefont {Duine},
  \citenamefont {Verheijen}, \citenamefont {Swagten},\ and\ \citenamefont
  {Koopmans}}]{Lavrijsen2015}%
  \BibitemOpen
  \bibfield  {author} {\bibinfo {author} {\bibfnamefont {R.}~\bibnamefont
  {Lavrijsen}}, \bibinfo {author} {\bibfnamefont {D.~M.}\ \bibnamefont
  {Hartmann}}, \bibinfo {author} {\bibfnamefont {A.}~\bibnamefont {{Van Den
  Brink}}}, \bibinfo {author} {\bibfnamefont {Y.}~\bibnamefont {Yin}}, \bibinfo
  {author} {\bibfnamefont {B.}~\bibnamefont {Barcones}}, \bibinfo {author}
  {\bibfnamefont {R.~A.}\ \bibnamefont {Duine}}, \bibinfo {author}
  {\bibfnamefont {M.~A.}\ \bibnamefont {Verheijen}}, \bibinfo {author}
  {\bibfnamefont {H.~J.}\ \bibnamefont {Swagten}}, \ and\ \bibinfo {author}
  {\bibfnamefont {B.}~\bibnamefont {Koopmans}},\ }\href {\doibase
  10.1103/PhysRevB.91.104414} {\bibfield  {journal} {\bibinfo  {journal} {Phys.
  Rev. B}\ }\textbf {\bibinfo {volume} {91}},\ \bibinfo {pages} {104414}
  (\bibinfo {year} {2015})}\BibitemShut {NoStop}%
\bibitem [{\citenamefont {Yang}\ \emph {et~al.}(2015)\citenamefont {Yang},
  \citenamefont {Thiaville}, \citenamefont {Rohart}, \citenamefont {Fert},\
  and\ \citenamefont {Chshiev}}]{Yang2014}%
  \BibitemOpen
  \bibfield  {author} {\bibinfo {author} {\bibfnamefont {H.}~\bibnamefont
  {Yang}}, \bibinfo {author} {\bibfnamefont {A.}~\bibnamefont {Thiaville}},
  \bibinfo {author} {\bibfnamefont {S.}~\bibnamefont {Rohart}}, \bibinfo
  {author} {\bibfnamefont {A.}~\bibnamefont {Fert}}, \ and\ \bibinfo {author}
  {\bibfnamefont {M.}~\bibnamefont {Chshiev}},\ }\href {\doibase
  10.1103/PhysRevLett.115.267210} {\bibfield  {journal} {\bibinfo  {journal}
  {Phys. Rev. Lett.}\ }\textbf {\bibinfo {volume} {115}},\ \bibinfo {pages}
  {267210} (\bibinfo {year} {2015})}\BibitemShut {NoStop}%
\end{thebibliography}%

\end{document}